\documentclass[11pt,twocolumn]{article}

\usepackage[utf8]{inputenc}
\usepackage[T1]{fontenc}
\usepackage[margin=1in]{geometry}
\usepackage{amsmath}
\usepackage{amssymb}
\usepackage{graphicx}
\usepackage{array}
\usepackage{longtable}
\usepackage{booktabs}
\usepackage{float}
\usepackage{caption}
\usepackage[hidelinks]{hyperref}

\newcommand{\keywords}[1]{%
    \par\medskip
    \noindent\textbf{Keywords:} #1
    \par\medskip
}

\hypersetup{%
    pdfauthor={Noah J. Bagazinski, Md. Ferdous Alam, Jaya Manideep Rebbagondla, Faez Ahmed},
    pdftitle={MiDShip: Multimodal Dataset of Ship Cargo Hold Structures for Engineering Design},
    pdfkeywords={Ship Design, Naval Architecture, Machine Learning, Structural Design, Engineering Dataset}
}

\begin{document}

\author{%
    Noah J.~Bagazinski\textsuperscript{1,*} \quad
    Md. Ferdous~Alam\textsuperscript{2}\\[0.4em]
    Jaya~Manideep~\mbox{Rebbagondla}\textsuperscript{1} \quad
    Faez~Ahmed\textsuperscript{1}\\[0.75em]
    \small \textsuperscript{1}Department of Mechanical Engineering,
    Massachusetts Institute of Technology\\
    \small Cambridge, MA 02139\\
    \small \texttt{noahbagz@mit.edu};
    \texttt{manideep@mit.edu};
    \texttt{faez@mit.edu}\\[0.5em]
    \small \textsuperscript{2}George W. Woodruff School of Mechanical Engineering,
    Georgia Institute of Technology\\
    \small Atlanta, GA 30318\\
    \small \texttt{ferdous@gatech.edu}\\[0.5em]
    \small \textsuperscript{*}Corresponding author
}

\date{}

\title{MiDShip: Multimodal Dataset of Ship Cargo Hold Structures for Engineering Design}

% keywords can be removed
\maketitle

\begin{abstract}
Ship structures govern a vessel’s strength, safety, and manufacturability. Their design is governed by classification-society rules comprising hundreds of requirements, making the design process complex and highly iterative. However, data-driven structural design is limited by the lack of large, structured datasets that connect design geometry with structural geometric measures and rule-based constraints. This paper presents MiDShip, a multimodal dataset of 12,753 synthetically generated cargo-hold structural designs for machine learning and optimization. It contains 6,020 randomly generated designs, 496 designs produced by a stochastic gradient Langevin dynamics (SGLD)-inspired procedure, and 6,237 designs produced by an equation-informed repair procedure. Each retained design includes a parametric vector, full 3D geometries and simplified geometries formatted for finite-element analysis (FEA) meshing, engineering drawings and corresponding annotated versions, a bill of materials, and structural geometric measures. The dataset also includes twenty-five constraints formulated from classification society rules for stiffener spacing, thickness, and section modulus requirements. Full-vessel loading, buckling, fatigue, and ice-class checks are outside the present scope. The primary contribution is the dataset and codebase; two design-generation studies demonstrate their use rather than introduce new optimization algorithms. MiDShip supports future research in machine learning, generative design, and automated rule-based evaluation for ship structures.
\end{abstract}

\keywords{Ship Design, Naval Architecture, Machine Learning, Structural Design, Engineering Dataset}

%%%%%%%%%%%%%% INTRO
\begin{figure*}[t]
\centering\includegraphics[width=\textwidth]{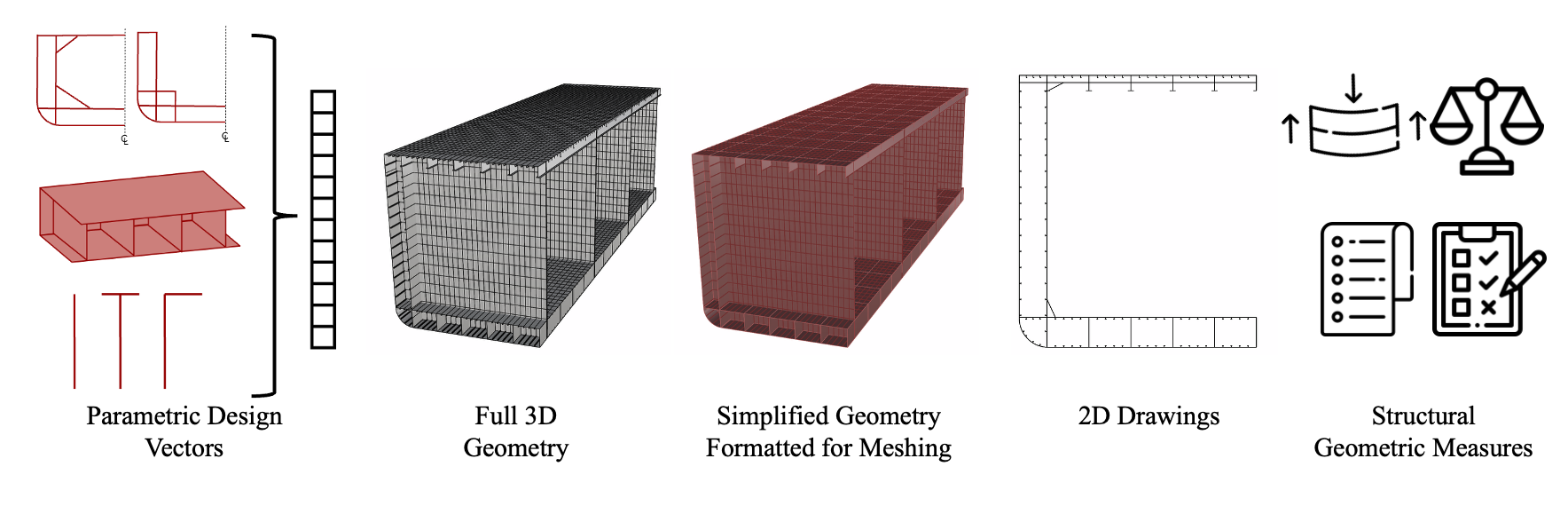}
\caption{Overview of the MiDShip dataset, a multimodal dataset for ship cargo-hold structural design and evaluation. Each design is represented by a parametric design vector used to generate full 3D geometries and simplified geometries formatted for finite-element analysis (FEA) meshing. Corresponding 2D engineering drawings are produced alongside labels for structural density, longitudinal cross-sectional area, center of gravity, maximum bending moment, and twenty-five local scantling evaluations derived from ABS design rules.}
\label{fig_1}
\end{figure*}

\section{Introduction}
Ship structural layout influences cargo-space arrangement, strength, manufacturability, lifecycle cost, and operational efficiency. These decisions affect a ship's steel weight, cargo capacity, fuel consumption, and emissions. This paper focuses on the design of a ship's parallel midbody structure, the constant-cross-section region containing the primary cargo spaces and central to longitudinal-strength evaluation. Cargo-hold structures are governed by individual classification-society rules. The International Association of Classification Societies (IACS) develops unified requirements and Common Structural Rules for adoption by its member societies, whose rules cover more than 90\% of the world's cargo-carrying tonnage~\cite{IACS_CSR_2024}. Detailed classification rules can form part of a goal-based standards framework when verified against its goals and functional requirements~\cite{IMO_GBS_2019}. As structural rules evolve, shared design representations and evaluation tools could help researchers examine the consequences of proposed requirements.

Digital engineering and artificial intelligence can accelerate design iteration and tradeoff analysis across marine engineering~\cite{papanikolaou2024ship, alam2024automation, picard2025concept}. Open-source tools such as \emph{Ship Workbench} and \emph{Vessel.js} provide extensible ship-evaluation modules~\cite{FreeCAD_ship, gaspar2018vessel,gaspar2022current}, while OOFEM provides an accessible structural-analysis framework~\cite{legaz2025OpenSource, palaversa2020oofem}. However, machine-learning research in ship structural design is constrained by limited access to shared design data. Shipyards and design firms generally retain their structural designs privately, and individual collections may contain too few examples for particular learning tasks. Public datasets linking design parameters, geometry, drawings, structural geometric measures, and rule evaluations would provide common data for developing and comparing methods.

To address this gap, this paper presents MiDShip, a multimodal dataset of ship cargo-hold structural designs generated by random sampling, an SGLD-inspired procedure, and an equation-informed repair procedure. The SGLD-generated and repaired subsets were derived from selected random-design seeds and therefore are derivative subsets rather than statistically independent samples. Each retained design includes a parametric design vector, full 3D geometries, and simplified geometries formatted for FEA meshing. It also includes three two-dimensional engineering drawings, an annotated version of each drawing, and a bill of materials. Evaluated outputs comprise structural geometric measures and twenty-five constraint evaluations derived from American Bureau of Shipping (ABS) design rules. An accompanying codebase was developed using the Python scripting tools for \emph{Rhinoceros 3D}. Figure~\ref{fig_1} provides an overview of the dataset and its components.

MiDShip is intended for researchers, naval architects, and shipyards developing data-driven methods for ship structural design. Its common parametric representation of tanker, containership, and bulk-carrier cargo-hold structures provides a starting point for benchmarking compatible designs against a public set. The paired design representations could support drawing-to-CAD reconstruction, drawing-based structural evaluation, and other drawing-based engineering tasks. Models could also be initially trained on MiDShip and then fine-tuned using a firm's private data. These applications are opportunities for future evaluation and are not benchmarked in this study.

The present study demonstrates the dataset workflow using selected requirements derived from existing ABS rules. Ice-class requirements and climate-dependent loading are outside the present evaluation scope. Section~\ref{subsec:dataset-limitations} describes the geometric assumptions and evaluation limits. Future extensions for studying the effects of revised rules and supporting rule-parameter calibration are discussed in Section~\ref{subsec:future-work}.

The key contributions of this work are:

\begin{enumerate}

\item A publicly released multimodal dataset of 12,753 cargo-hold structural designs, of which 5,274 satisfy all twenty-five implemented constraints. Each design includes synchronized parametric, geometric, drawing, and material representations, structural geometric measures, and rule-based evaluations. The dataset is released under the GNU General Public License version 3.0.
\item An accompanying parametric CAD and rule-based evaluation pipeline for generating, modifying, and evaluating cargo-hold structures. The accompanying software is released under the GNU General Public License version 3.0.
\item A preliminary demonstration that surrogate models can learn relationships among design parameters, structural geometric measures, and the twenty-five implemented constraints.
\item Two demonstrations of constraint-aware data-driven design generation using the information in the MiDShip Dataset: a surrogate-guided, SGLD-inspired procedure and an equation-informed parametric-repair procedure.
\end{enumerate}

%%%%%%%%%%%%%%%%% PRIOR WORK
\section{Prior Work}
Relevant prior work spans ship structural evaluation, engineering datasets, AI-assisted ship design, and automated drawing interpretation.

\subsection{Ship Structure Evaluation}

Prior work in ship structural evaluation has primarily focused on finite element methods, with the overarching goal of ensuring compliance with classification society rules such as the \emph{IACS Common Structural Rules}. Within this framework, researchers have explored parameterized structural design and optimization approaches. For example, reduced-order modeling has been used to represent cruise ship structures and enable Bayesian optimization to minimize deflection, yielding, and buckling of structural components~\cite{fabris2025data}. Similarly, longitudinal ship structures have been optimized for weight and cost using FEA combined with NSGA-2, while enforcing regulatory constraints defined by Bureau Veritas~\cite{abedin2024two}. Several studies have also evaluated representative three-cargo-hold configurations as benchmark problems for structural analysis and validation~\cite{palaversa2020oofem, fabris2025data, gkougkoudis2022finite, servis2003finite}. 

Parametric modeling and rule-based sizing of ship structures predate the present work. Varela et al. developed a hull-structure product data model and digital-prototyping system for rapid parametric generation of three-dimensional structural models at the basic-design stage~\cite{varela2011product}. Ryumin and Tryaskin presented CADS-Hull, a computer-aided system for parametric hull-structure design under classification-society requirements~\cite{ryumin2022cadsHull}. These systems establish parametric structural modeling and rule-informed design as prior capabilities; MiDShip's contribution is instead the release of a reusable, machine-learning-oriented dataset that synchronizes parametric designs with CAD geometry, engineering drawings, bills of materials, structural properties, and evaluations of the twenty-five implemented constraints.

Other studies have examined the complexity of constraints in ship structural design and developed data-driven methods to repair designs to satisfy them. Two studies developed adaptive repair methods that identify the design variables associated with each violated constraint and replace them with values from feasible or minimally violating donor designs, improving the convergence and diversity of ship structure designs in optimization tools such as NSGA-2 and MOEA/D~\cite{samanipour2020adaptive,jelovica2024improved}. Cai and Jelovica subsequently used neural networks to discover variable--constraint mappings for autonomous repair in multi-objective structural optimization, including a chemical-tanker case~\cite{cai2023neural}. Related work on concept-stage structural optimization includes Raikunen et al., who combined finite-element modeling, particle swarm optimization, and indirect stress-constraint relaxation to reduce passenger-ship structural steel weight~\cite{Raikunen03102019}.

Despite these advances, reusable public benchmark data and accessible codebases for data-driven ship structural design remain limited. This work aims to provide tools for future ship structure researchers and designers.

\subsection{Engineering Design Datasets}

Engineering-design machine learning requires datasets that connect structured design information with performance labels. Useful datasets should be domain-relevant, sufficiently diverse, practically applicable, interoperable, well documented, and explicit about design constraints~\cite{ahmed2025design}.

Many engineering datasets are synthetically generated through procedural design scripts and evaluated using engineering simulations. When constructing such datasets, important considerations include compatibility of the design representation with common engineering software, completeness of captured design features, data compactness, and generalizability to multiple design tasks. Additional decisions must be made regarding which performance objectives to include, overall dataset size, and sampling strategy~\cite{picard2023dated}.

Publicly available synthetic datasets for mobility-related applications include bicycle frames~\cite{regenwetter2022framed, regenwetter2022biked, regenwetter2024bikedpp}, automotive components~\cite{yoo2026deepwheel, wollstadt2022carhoods10k}, vehicle exterior design~\cite{elrefaie2024drivaernet, elrefaie2024drivaernetpp}, blended wing body aircraft~\cite{sung2025blendednet}, and ship hulls~\cite{bagazinski2023shipD, read2009drag}. These works demonstrate the growing role of synthetic, simulation-driven datasets in data-driven engineering design.

Among the closest public engineering datasets, Ship-D provides parametric external hull forms with geometric, hydrostatic, and hydrodynamic labels~\cite{bagazinski2023shipD}, whereas BIKED provides parametric bicycle designs, images, component representations, class labels, and machine-learning benchmarks~\cite{regenwetter2022biked}. MiDShip differs in both design object and labeling: it focuses on internal cargo-hold structures and synchronizes rule-derived structural evaluations with parametric, CAD, drawing, bill-of-materials representations and structural geometric measures. Thus, MiDShip applies the reusable engineering-dataset paradigm established by these resources to classification-rule-governed ship structural design.

\subsection{Artificial Intelligence for Ship Design}

Most prior work in artificial intelligence for ship design has focused on hull form development, where the primary objective is to minimize hydrodynamic resistance while maintaining sufficient displacement for operational requirements. High-fidelity hydrodynamic evaluation methods, such as computational fluid dynamics, are computationally expensive, motivating the use of AI to accelerate performance prediction~\cite{htein2025ai, papanikolaou2024ship}. A common approach involves parametric design, where hull geometries are generated using predefined functions and paired with simulated performance data to train surrogate models. These models can reduce evaluation time by orders of magnitude and enable rapid optimization and generative design workflows~\cite{bagazinski2023shipD, bagazinski2023shipgen, read2009drag, yonekura2023designing}. Alternative approaches learn latent representations of hull geometries directly from 3D data or employ free-form deformation techniques to modify and reconstruct hull shapes, supporting efficient dataset generation and design exploration~\cite{hodgesai, khan2023shiphullgan, khan2022shape, wang2022shipEncoding, ao2021artificial, ao2022artificial}.

Surrogate-guided Langevin sampling has also been applied to multi-constraint engineering inverse design using learned gradients for expensive black-box evaluations~\cite{nguyen2020nanomaterial}; the present work applies a related SGLD-inspired concept to ship structures.

However, a ship is defined not solely by its hull but by an integrated system composed of structural, spatial, and functional components. Some studies have extended computational design methods to broader ship-system considerations, including general arrangement design and layout optimization~\cite{kim2025review, le2022fly, HybridAgent, ShipArrangements}. Additionally, system-level optimization frameworks have been explored across varying levels of design fidelity, incorporating multidisciplinary considerations such as performance, cost, and operational constraints~\cite{papanikolaou2010holistic, hart2010IMDO, solheim2025system}. Despite these advances, the majority of AI-driven ship design research remains concentrated on hull optimization, with comparatively limited attention given to structural design and associated data-driven methodologies.

\subsection{Drawing Evaluation and Design Verification with Artificial Intelligence}

Earlier ship-specific work mapped information from 2D midship drawings into STEP AP218 three-dimensional structural models and recognized design symbols for automated information extraction~\cite{hwang2004mapping,hwang2005recognition}. MiDShip instead provides paired drawings, annotations, and CAD representations as dataset modalities.

Recent advances in computer vision and machine learning have demonstrated the feasibility of automated interpretation of engineering drawings. Early approaches focused on detecting graphical primitives and symbols, enabling the extraction of structured information from engineering documentation~\cite{lin2023integration, yang2024comprehensive}. Neural networks have shown increasing capability to identify geometric features, textual content, and semantic relationships in complex diagrams~\cite{jamieson2024review}. More recent work has improved robustness through fine-tuning on engineering drawing datasets, with pipelines developed to segment drawings, detect title blocks, and extract annotations~\cite{khan2025automated, khan2025multi}. Few-shot learning approaches have further improved symbol recognition with limited labeled data~\cite{jamieson2024few}, while automated pipelines have generated 2D drawings from 3D CAD geometries, creating datasets for design evaluation~\cite{wortham2025linesTo3D}. Additionally, vision-language models have been benchmarked on engineering information-extraction tasks, including applications to marine systems~\cite{khallouli2022leveraging, doris2025designqa, picard2025concept}. However, these advances depend critically on access to curated datasets combining drawings, design representations, and associated performance data.

This dataset requirement mirrors the broader infrastructural gap identified across ship design research. While hull form optimization has benefited from parametric design and simulation-driven datasets, structural design studies have remained largely isolated—confined to case-specific finite element models rather than accessible, reusable resources. Consequently, drawing-based evaluation and automated verification workflows for ship structures lack the foundational data infrastructure necessary to leverage AI effectively. A comprehensive dataset integrating parametric designs, engineering drawings, structural geometric measures, and rule-based evaluations would directly enable these AI-driven evaluation and verification workflows for ship structural design.

\subsection{Research Gap and Project Motivation}

Taken together, the prior literature highlights substantial progress in computational ship design, structural analysis, and engineering artificial intelligence, while also revealing a clear gap in available research infrastructure. Existing ship AI studies have focused primarily on hull form optimization, while structural design studies have largely relied on case-specific finite element models rather than reusable datasets. Similarly, recent advances in engineering drawing interpretation and multimodal learning demonstrate the growing value of structured design data, yet comparable benchmarks for ship structures remain scarce. To the authors' knowledge, based on the literature reviewed through August 2026, MiDShip is the first publicly released, machine-learning-oriented benchmark to synchronize parametric cargo-hold structural designs, CAD geometries, engineering drawings, bills of materials, structural geometric measures, and classification-rule constraint evaluations within a single resource. The MiDShip dataset is intended to address this gap and provide a foundation for future research in ship structural machine learning, generative design, and automated engineering evaluation.

%%%%%%%%%%%%%%%%%%%%%%%%%%%%%%%%%%%%%%%%%% DATASET DETAILS
\begin{figure}[h]
\centering\includegraphics[width=\columnwidth]{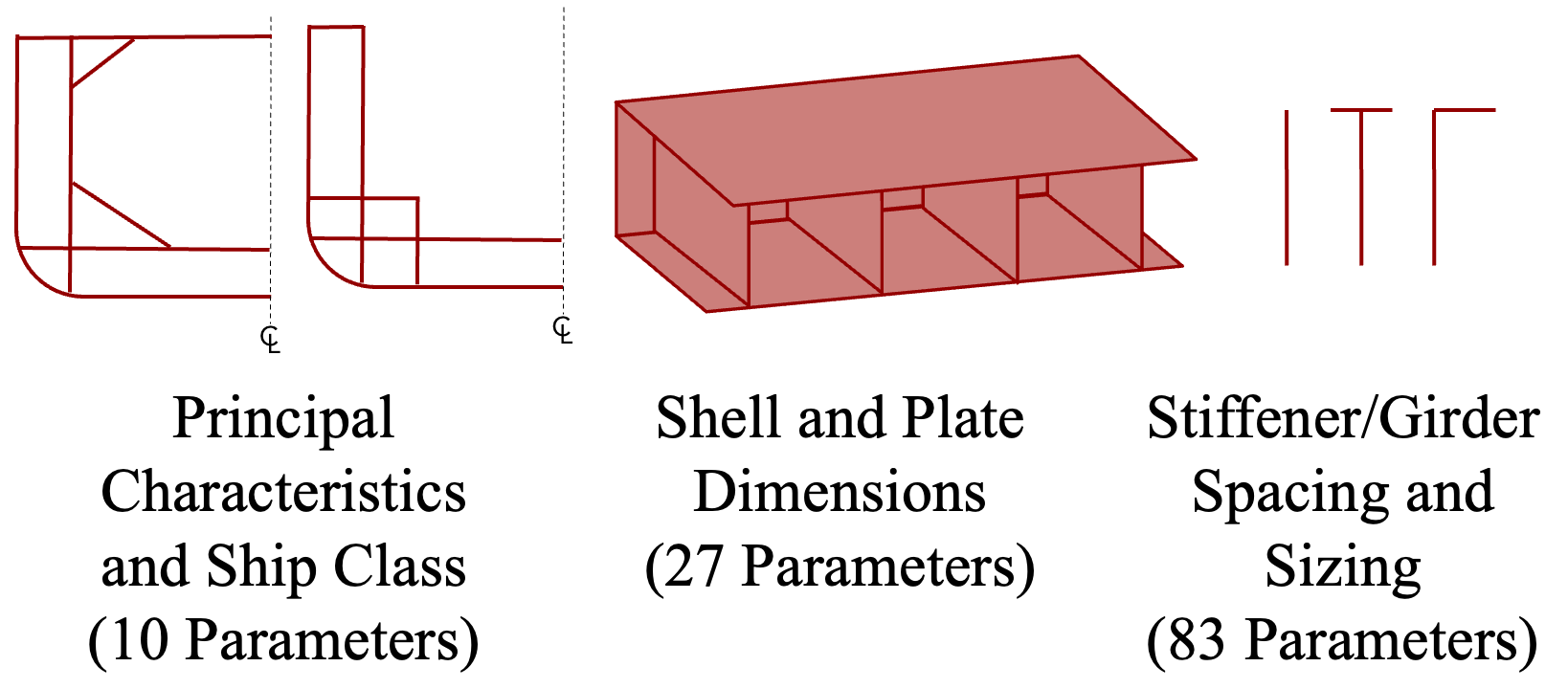}
\caption{Each ship structure is defined by a set of 120 parameters grouped into three categories: principal characteristics and ship class (10 parameters), shell and plate dimensions (27 parameters), and stiffener/girder spacing and sizing (83 parameters). There are twenty unique stiffeners characterized by the parametric designs.}
\label{fig_2}
\end{figure}

\section{Dataset Details}
MiDShip focuses on the cargo-hold structures of tankers, containerships, and bulk carriers. This region is central to longitudinal strength and cargo space design. Design rules for this portion of a ship structure are relatively standardized with classification society requirements~\cite{IACS_CSR_2024, ABS_MVR}. These characteristics provide a focused domain for data-driven structural design and evaluation.

Each retained design includes parametric data, full 3D geometries and simplified geometries formatted for FEA meshing, engineering drawings, a bill of materials, structural geometric measures, and constraint evaluations. The dataset comprises 6,020 random, 496 SGLD-generated, and 6,237 repaired designs. Because the latter two subsets derive from selected random-design seeds, the subsets are related rather than statistically independent. The synthetic random subset follows prescribed parameter ranges and coupling rules rather than an empirical fleet distribution. The following subsections describe the parametric designs, geometries, and drawings.

%\begin{table*}[t]
%\caption{Data modalities included for every retained design in each MiDShip subset. The drawing package contains three engineering drawings and three corresponding versions with labeled bounding boxes and associated metadata.}
%\label{tab:subset_modalities}
%\centering
%\resizebox{\textwidth}{!}{
%\begin{tabular}{lcccccc}
%\toprule
%Dataset subset & Parametric data & Full 3D geometry & Mesh-ready geometry & Drawing package & Bill of materials & Properties and constraints \\
%\midrule
%Randomly generated (6,020) & Included & Included & Included & Included & Included & Included \\
%SGLD-generated (496) & Included & Included & Included & Included & Included & Included \\
%Equation-informed repaired (6,237) & Included & Included & Included & Included & Included & Included \\
%\bottomrule
%\end{tabular}
%}
%\end{table*}

\subsection{Parametric Design}
\begin{figure*}[t]
\centering\includegraphics[width=\textwidth]{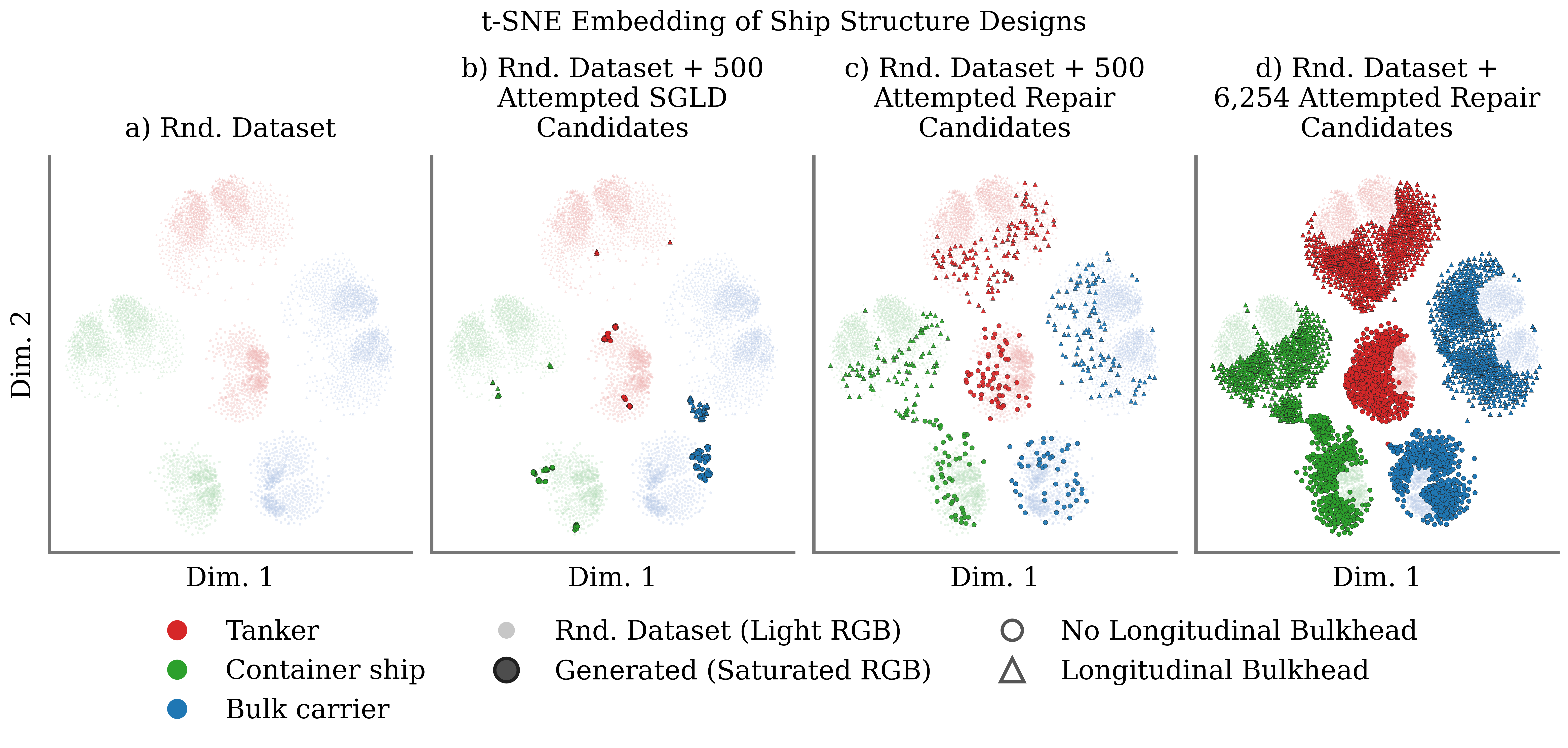}
\caption{Two-dimensional t-SNE visualizations of the design distributions, colored by ship class and the presence of longitudinal bulkheads. Panel (a) shows the 6,020 random designs; panel (b) overlays all 500 attempted SGLD candidate vectors; panel (c) overlays a 500-design sample of repaired candidate vectors; and panel (d) overlays all 6,254 attempted repair candidate vectors. The projection provides a qualitative view of the jointly embedded parameter vectors and is not used to quantify global design-space coverage or engineering diversity.}
\label{fig_3}
\end{figure*}

MiDShip geometries are generated from 120 parameter fields: 10 describe principal characteristics and ship class, 27 describe major plate structures, and 83 describe girder and stiffener dimensions and spacing. These fields control global geometry and local structural details while providing a consistent tabular representation for optimization and machine learning. Figure~\ref{fig_2} highlights the represented structural elements. The Design Parameters appendix lists all 120 parameter fields, their units, and nominal sampling specifications in Table~\ref{tab_4}.

All designs use evenly spaced transverse bulkheads; a solid double-bottom deck with full-height floor beams and longitudinal girders; and regularly spaced stiffeners and girders. When present, longitudinal bulkheads divide the hold into halves, thirds, or quarters across the beam. Real ships, however, may use localized spacing based on loading. These assumptions simplify the design space while preserving dominant cargo-hold structural characteristics; further discussion appears in the Limitations Section.

The first generated set in the MiDShip Dataset was randomly sampled from the parametric design space. Random inputs were uniformly sampled within the nominal ranges and coupled relationships listed in Table~\ref{tab_4}. Integer fields were sampled as integers, and plate thicknesses were rounded to whole millimeters, although shipyards may use plate catalogs. The stored $T$ field was sampled as a fraction of $D$ but was unused; rule calculations instead used $T_{\mathrm{rule}}=0.667D$. Ship class was one-hot encoded, and class-specific fields were activated only for the selected class and otherwise set to zero. Bracket dimensions were subsequently aligned with neighboring members. The codebase includes functions to automatically clean parametric vectors based on the class and automatically align brackets to land at the locations of stiffeners or girders.

Using the parameters, two methods were used to characterize and visualize the different generated design sets. The first method used the mean nearest-neighbor distance (MNN) for equal-size sets of $n=100$ designs or candidate vectors. For each 100-design set, MNN averages the distance from each design to its nearest neighbor in the scaled parameter space. Each of the 120 parameters was min--max scaled using the observed minima and maxima of the complete retained random subset; constant columns were set to zero, and scaled values were clipped to $[0,1]$. For scaled parameter vectors $\mathbf{z}_i$, MNN was calculated as
\begin{equation}
    \operatorname{MNN}=\frac{1}{n}\sum_{i=1}^{n}\min_{k\ne i}\left\|\mathbf{z}_i-\mathbf{z}_k\right\|_2.
\end{equation}
The random and repaired results use thirty sets sampled without replacement within each 100-design set using random seed 41; sets could overlap across repeated draws. The SGLD result uses the five actual 100-candidate generation batches. MNN values are reported as the mean $\pm 2s$ across these equal-size sets, where $s$ is the sample standard deviation. These ranges describe between-set dispersion and are not confidence intervals. The MNN across the sample sets is provided in Table~\ref{tab_1}.

In addition to the relative diversity of samples, a 2D t-SNE was made across all of the samples in the dataset for visualization. Figure~\ref{fig_3} provides a qualitative visualization of the relative parametric diversity of the random, SGLD-generated, and repaired samples. All vectors were embedded jointly after standardizing the 120 parameters and weighting the ship-class and longitudinal-bulkhead indicators by 2.5; PCA reduced the data to 40 components before t-SNE was applied with perplexity 45 and random seed 41. The visualization complements the quantitative MNN comparison by showing how the samples cluster within the parameter space. Because t-SNE does not preserve global distances, this visualization is not treated as a quantitative measure of design-space coverage.

This sampling framework enables large-scale generation of synthetic structural designs with controlled parametric variation. It does not establish global coverage of feasible ship structures or the empirical realism of the resulting distribution.

\begin{figure*}[t]
\centering\includegraphics[width=\textwidth]{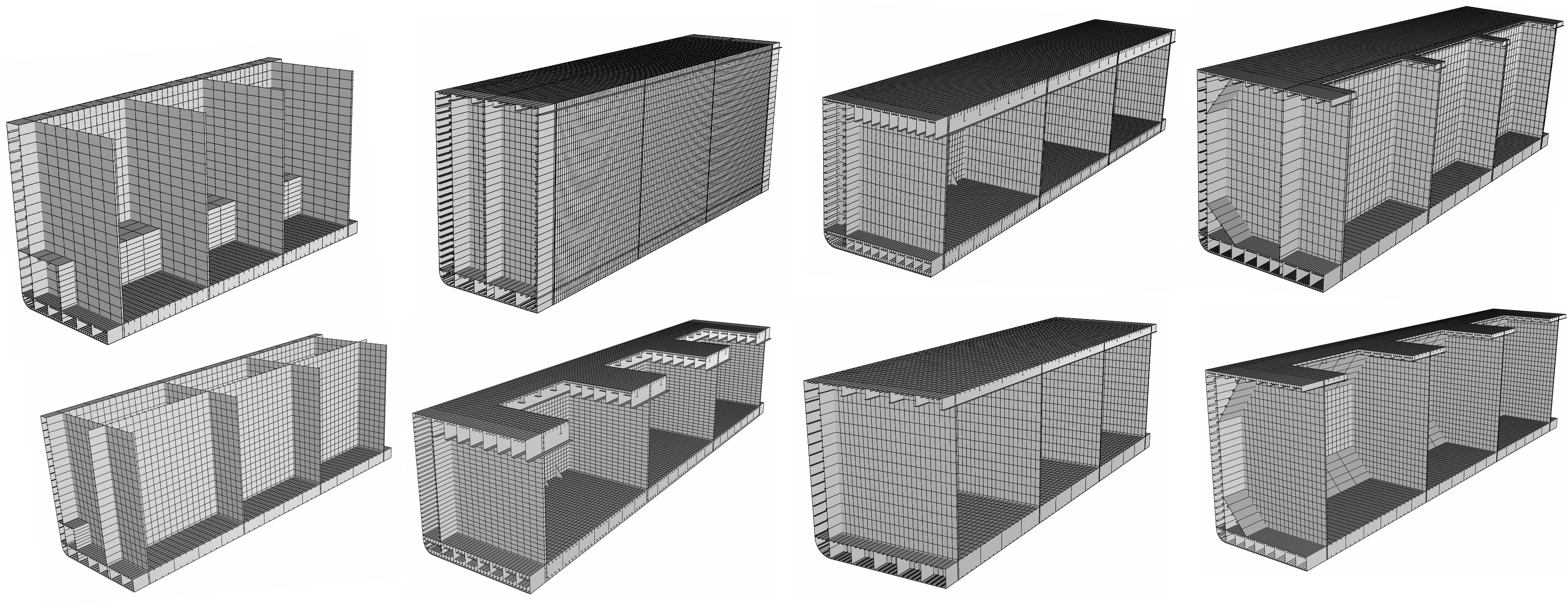}
\caption{Examples of dataset designs. The leftmost column shows two containership designs; the two middle columns show tanker designs with varying longitudinal bulkheads and hatch openings; and the rightmost column shows two bulk-carrier designs.}
\label{fig_4}
\end{figure*} 

\begin{figure}[h]
\centering\includegraphics[width=1.75in]{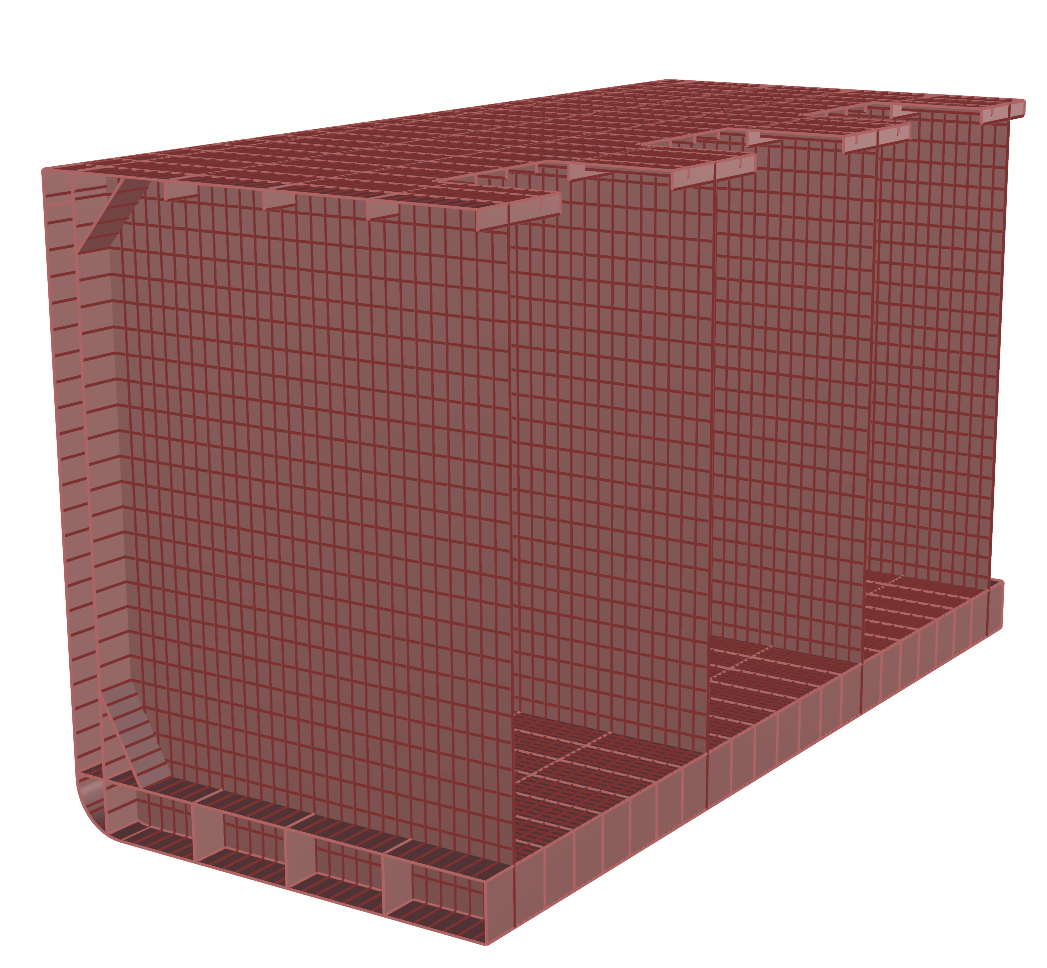}
\caption{Example simplified geometry formatted for FEA meshing. This representation is identical to the full 3D geometry except that smaller stiffeners are replaced with curves. The structural properties of each stiffener are stored in a list to support future finite-element meshing.}
\label{fig_5}
\end{figure} 
\subsection{Geometry Development}

Using the parametric framework, Python code was developed with the \emph{RhinoScriptSyntax} application programming interface in \emph{Rhinoceros 3D} (Version 8) to generate the 3D structural geometries. All geometries follow the standard ship coordinate system, in which the \textit{x}-axis is longitudinal, the \textit{y}-axis is transverse, and the \textit{z}-axis is vertical. The parameters define the hull shell, transverse bulkheads, decks, and longitudinal bulkheads. For each primary plate structure, stiffeners are then extruded along the principal axes of the plate. Except for the curved shell geometry at the bilge radius, all structural elements are planar.

As structural elements are procedurally generated in \emph{Rhinoceros 3D}, each component is assigned a globally unique identifier (GUID). For every element, the identifier, position, orientation, primary length, depth, thickness, planar area, cross-sectional area, and area moments of inertia are recorded. This metadata supports future structural analysis, CAD-based editing, traceability, and multimodal machine-learning applications. Figure~\ref{fig_4} provides examples of generated structures, illustrating the three ship classes as well as optional longitudinal bulkheads and hatch openings.

On average, a single design was generated in 3.25 seconds on a 2024 Apple M3 Pro processor with 12 cores using the \emph{Rhinoceros 3D} Python scripting interface. In total, 6,050 designs were attempted, and 30 parameter sets resulted in geometric errors, leaving 6,020 valid designs. This randomly generated portion of the dataset contains 1,934 containership designs, 2,006 tanker designs, and 2,080 bulk-carrier designs. This generation speed enables rapid expansion of the dataset and future large-scale design studies.

In addition to the full 3D geometries, a complete list of structural elements and their geometric properties is provided as a bill of materials. A simplified geometry formatted for FEA meshing was also created for each design. In this representation, shell plating, girders, web frames, and other major members retain their full three-dimensional geometry, while smaller stiffeners are represented by curves that can later be modeled as line elements in finite-element meshes. Unique identifiers for each structural element were embedded in both the full and simplified CAD files to preserve traceability between representations. Figure~\ref{fig_5} shows an example. This release does not include finite-element meshes or FEA results. A separate future study will describe the development and evaluation of automated meshing and finite-element analysis using these geometries.

\subsection{Engineering Drawing Development}
\begin{figure*}[t]
\centering\includegraphics[width=\textwidth]{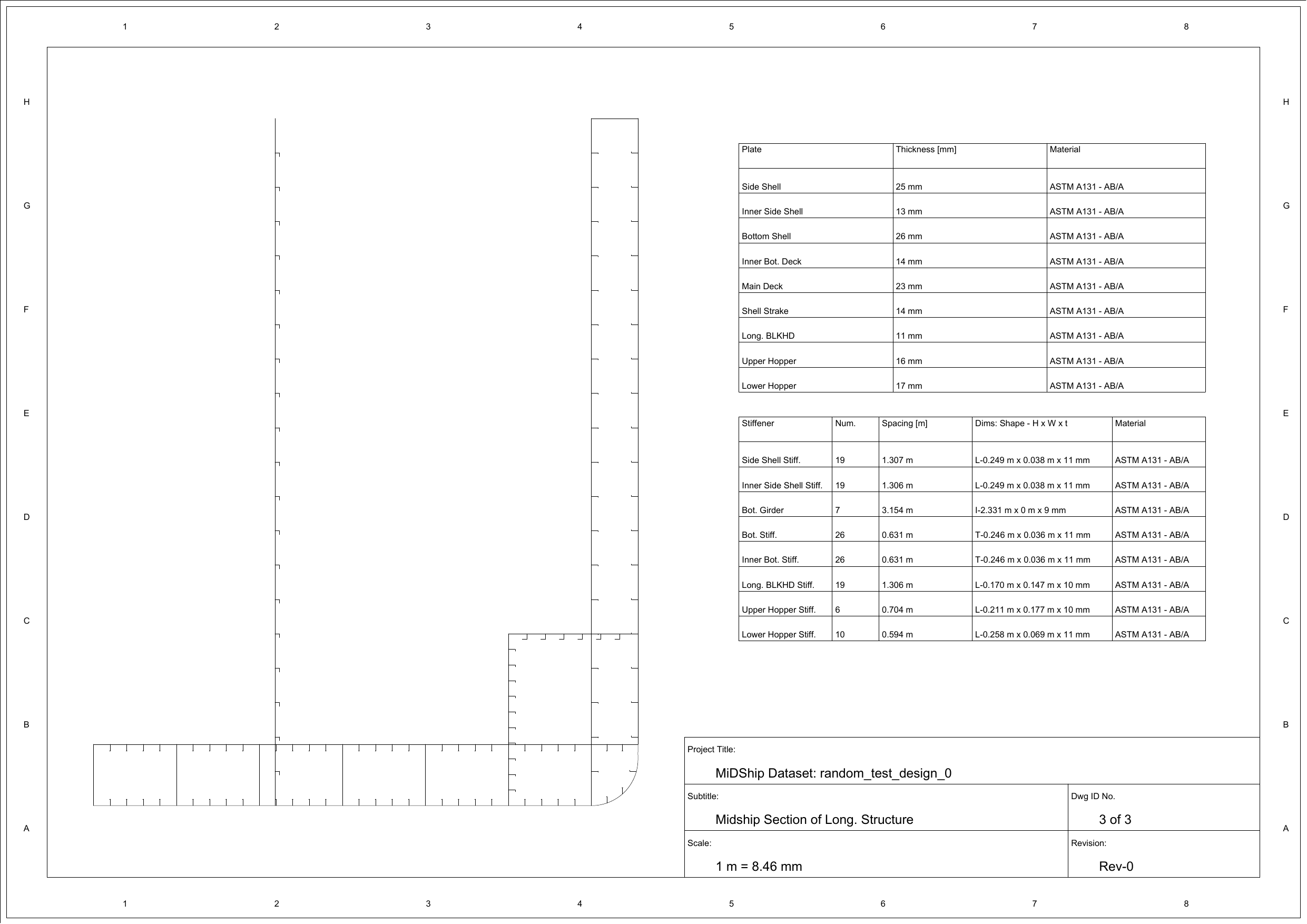}
\caption{Example 2D engineering drawing of the longitudinal structure. The drawing contains a slice of the ship structure, a title block, two information blocks, and formatted margins.}
\label{fig_6}
\end{figure*}

\begin{figure}[h]
\centering\includegraphics[width=\columnwidth]{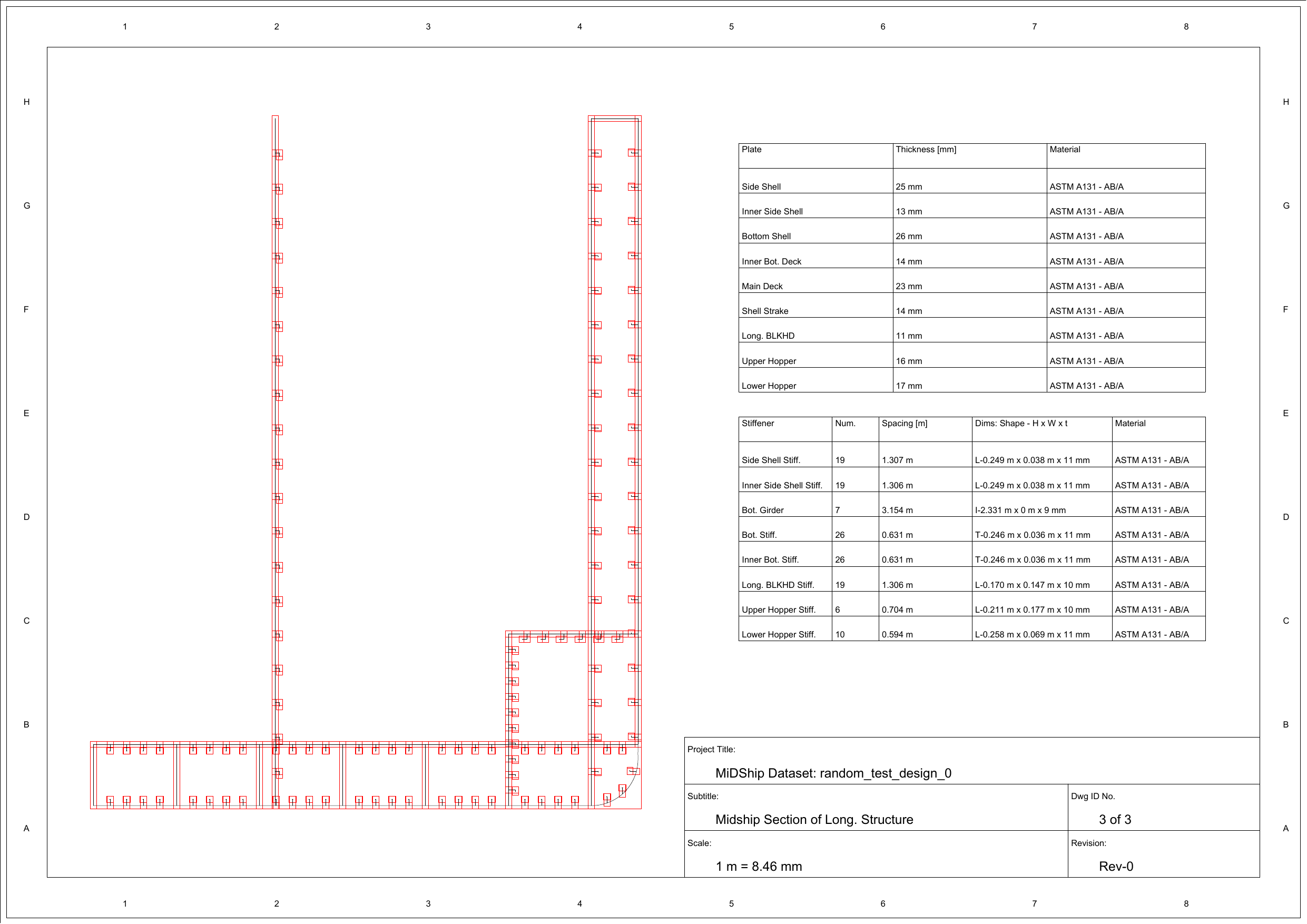}
\caption{The 2D drawing in Figure~\ref{fig_6} with labeled bounding boxes around structural elements. The positions, dimensions, and unique identifiers of the structural elements and bounding boxes are stored in a separate metadata file.}
\label{fig_7}
\end{figure}

After the structural geometries were generated, a set of three 2D engineering drawings was created for each design. These drawings correspond to the longitudinal structure, transverse bulkhead structure, and main webframe-beam-floor profiles. Each drawing includes the geometry, a title block, and two information blocks (plate data and stiffener data).
A second set of three drawings was also generated with labeled bounding boxes around structural components. Separate metadata files record the position, size, and scale of every structural element and its associated bounding box. Each entry also includes a unique identifier that enables traceable editing of individual components within the \emph{Rhinoceros 3D} scripting environment. These drawings were procedurally generated using Python tools in \emph{Rhinoceros 3D} through a procedure adapted from a previously developed automated drawing pipeline~\cite{wortham2025linesTo3D}.
Figure~\ref{fig_6} shows a longitudinal structural drawing with the design identifier, drawing type, scale, and member dimensions and spacing; Figure~\ref{fig_7} adds labeled bounding boxes. Although not benchmarked here, these drawings support future 2D-to-3D reconstruction, automated drawing generation, engineering-information extraction, and multimodal design reasoning.

\begin{figure*}[t]
\centering\includegraphics[width=\textwidth]{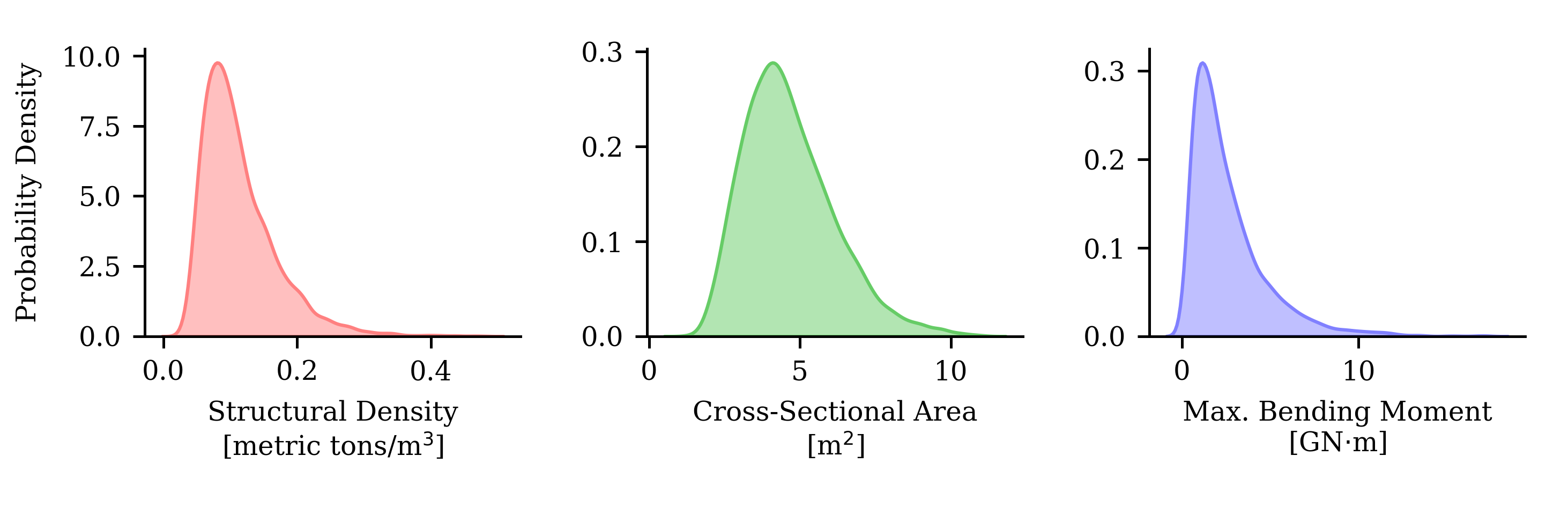}
\caption{Distributions of structural geometric measures for the 6,020 retained random designs. The plots show the volume-normalized structural steel mass, longitudinal cross-sectional area, and calculated maximum bending moment released with the dataset. Probability density is expressed in reciprocal units of the corresponding horizontal-axis quantity.}
\label{fig_8}
\end{figure*} 
\begin{figure}[h]
\centering\includegraphics[width=\columnwidth]{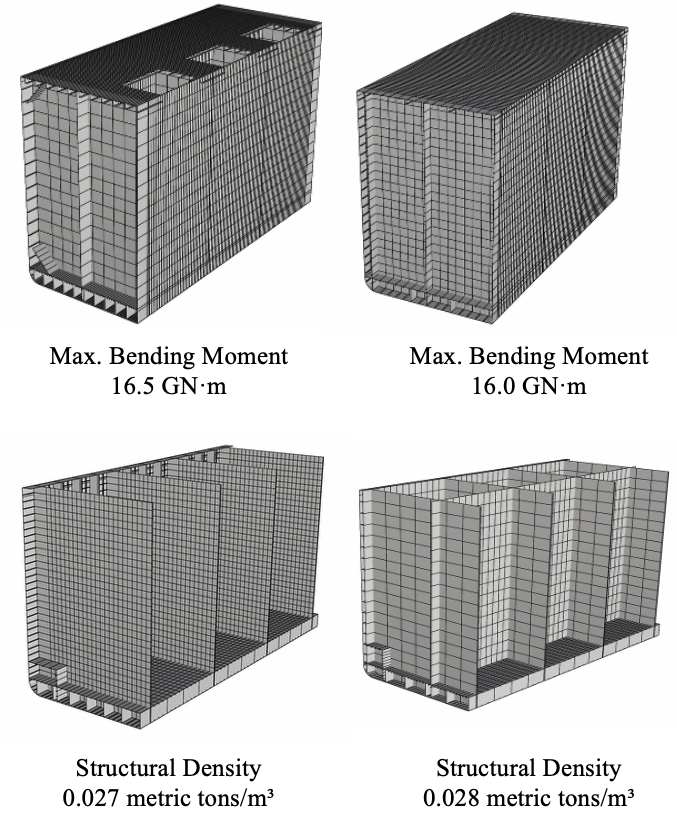}
\caption{Examples of random designs with high calculated maximum bending moments (top) and low structural densities (bottom).}
\label{fig_9}
\end{figure}

\subsection{Structural Geometric Measures}
Three structural geometric measures were calculated for surrogate modeling and design studies. These measures describe qualities of interest to designers: material use, effective longitudinal area, and a simplified estimate of longitudinal bending capacity. They are derived from geometry and the stated material assumptions, rather than from an analysis of structural response to applied loads.
The first measure is the volume-normalized structural steel mass, hereafter referred to as \emph{structural density}. Lightweight structural designs are desirable because they reduce material cost and allow greater cargo capacity for a given vessel displacement. All modeled structural members contribute to this measure. The generator represents structural members as plate elements. Each plate's surface area $A_{\mathrm{plate},i}$ and thickness $t_{\mathrm{plate},i}$ determine its volume, which is multiplied by steel density to calculate its mass. These elements include the shell, decks, bulkheads, girders, frames, and the web and flange plates of stiffeners. The generated structural members do not intersect within the explicitly modeled half-structure. Equation~\ref{eq_2} normalizes this mass by the rectangular volume defined by the three-hold length $L_{3h}$, beam $B$, and depth $D$. Throughout this paper, ``ton'' denotes a metric ton of 1,000~kg. Accordingly, the steel density $\rho_{\mathrm{steel}}$ was taken as 7.85~metric tons/m$^3$ (7,850~kg/m$^3$), and $N_{\mathrm{plate}}$ is the number of modeled plate elements. The factor of 2 accounts for port--starboard symmetry because only half of the structure is explicitly modeled. Under this convention, members located exactly on the centreline are also counted twice. In the mass calculation, this treats the centreline girder as having twice its specified thickness. This measure has units of mass per bounding volume and should not be interpreted as the material density of steel.

\begin{equation}
\label{eq_2}
    \rho_{\mathrm{struct}} = \frac{2\rho_{\mathrm{steel}}\sum_{i=1}^{N_{\mathrm{plate}}} A_{\mathrm{plate},i}t_{\mathrm{plate},i}}{L_{3h}BD}
\end{equation}

The second measure is the cross-sectional area of the effective longitudinal structure. Only structural elements oriented in the longitudinal direction and spanning the modeled three-hold length contribute to this calculation. In the bill of materials and associated metadata, these members are identified by \texttt{x\_dir}=1 and the full modeled length. Equation~\ref{eq_3} provides the calculation using the stored cross-sectional area $A_{\mathrm{cx},i}$ of each contributing plate element. For a stiffener, the summed elements include the cross-sectional areas of both its web and flange.

\begin{equation}
    \label{eq_3}
    A_{\mathrm{cx}} = 2\sum_{i=1}^{N_{\mathrm{long}}} A_{\mathrm{cx},i}
\end{equation}

The third measure is the calculated maximum bending moment, $M_{\max}$. This calculation is a simplified yield-based elastic proxy for longitudinal bending capacity. Ship structures experience longitudinal bending loads arising from buoyancy, weight, and waves, and their elastic section properties provide one indication of longitudinal capacity. Equation~\ref{eq_4} shows the nominal formulation from the geometry of the effective longitudinal elements.

\begin{equation}
    \label{eq_4}
    M_{\max} = \frac{0.5\sigma_y I_{\mathrm{struct}}}{z_{\max}}
\end{equation}

In this equation, the assumed steel yield strength $\sigma_y$ is 250~MPa, $I_{\mathrm{struct}}$ is the second moment of area of the effective longitudinal structure, and $z_{\max}$ is the maximum vertical distance from the neutral axis. The factor 0.5 applies a safety factor of two selected by the authors for this dataset. This is a user-selected research parameter rather than a rule-derived value; another safety factor could be substituted when recalculating the measure for a different application. The maximum bending moment calculation is based on section properties and yield strength; it is not a wave-loading demand, ultimate-strength calculation, or regulatory allowable bending moment. It does not account for buckling, nonlinear response, fatigue, load distribution, or other failure modes. Equation~\ref{eq_5} gives the nominal effective second moment of area using the parallel-axis theorem.

\begin{equation}
    \label{eq_5}
    I_{\mathrm{struct}} = 2\sum_{i=1}^{N_{\mathrm{long}}}\left[I_{y,i} + \left(z_i-z_{\mathrm{NA}}\right)^2 A_{\mathrm{cx},i}\right]
\end{equation}

In the calculation, $N_{\mathrm{long}}$ is the number of longitudinal elements spanning the three-hold structure, $I_{y,i}$ is the local second moment of area of element $i$, $z_i$ is its vertical centroid, $z_{\mathrm{NA}}$ is the neutral-axis height of the effective longitudinal structure, and $A_{\mathrm{cx},i}$ is its cross-sectional area. For stiffeners, the web and flange are both included as contributing elements; their generated T- or C-section arrangement is reflected through the element centroids and their parallel-axis contributions to $I_{\mathrm{struct}}$.

Figure~\ref{fig_8} shows the distributions of structural density, cross-sectional area, and maximum bending moment in the retained random subset. Figure~\ref{fig_9} shows examples with high maximum bending moments and low structural densities.

\subsection{Design Rule-Based Constraints}

A set of twenty-five design constraints was selected to evaluate the scantling structural systems. These requirements are derived from the July 2024 ABS \emph{Marine Vessel Rules} and apply to a selection of structural elements. The selected constraints can be evaluated without a separate vessel-specific loading analysis. Their threshold calculations use geometric quantities and prescribed loading assumptions within the implemented rule formulations~\cite{ABS_MVR}. The rules define minimum thicknesses, maximum spacings, and required section moduli for structural members. The evaluator consolidates 38 rule-to-output applications drawn from 22 cited ABS clauses into twenty-five constraint outputs. When multiple applicable minimum requirements contribute to one output, the greatest, and therefore strictest, required minimum is used. The bottom-floor-spacing output is the sole maximum-type constraint and uses its applicable maximum-spacing requirement. The accompanying codebase provides the implemented equations, units, aggregation rules, and ship-class applicability logic for all twenty-five outputs; for each design, the evaluator applies the requirements relevant to the encoded ship class and records the governing threshold. Table~\ref{tab_5} in the Appendix lists these rule applications and their associated constraint outputs. The stored $T$ field was not used as a rule input; calculations requiring draft used $T_{\mathrm{rule}}=0.667D$. Some rule calculations also require the total ship length. In these cases, a reference length of 250 meters was used because full-vessel particulars are not known a priori for every generated design.

For each constraint, formulations based on ABS Rules were used to compute allowable threshold values from the design parameters. These thresholds were then compared with geometric properties extracted from the 3D models, including plate thicknesses, section-modulus values, and member sizes. This set of constraints does not represent the complete list of structural evaluations required for full ship approval. However, the constraints provide a rule-informed basis for evaluating selected scantling requirements in a ship's cargo-hold region. Additional ABS requirements include global bending loads from cargo and buoyancy, slenderness ratios, longitudinal-member spacing, and bulkhead-spacing requirements. These requirements were excluded because they require full-vessel information beyond the three-hold structural region. This limitation is discussed further in the Limitations Section.

Throughout this paper, a \emph{feasible design} is one that satisfies all twenty-five constraints formulated and implemented in this work. This term does not imply satisfaction of all requirements for complete ship structural design or ABS approval.

Let $v_j(\mathbf{x})$ denote the signed geometric value for constraint $j$, and let $t_j(\mathbf{x})$ denote its signed rule threshold. The twenty-four minimum requirements retain their physical signs. For the sole maximum requirement, bottom-floor spacing, both the physical spacing and its maximum allowable value are multiplied by $-1$. Consequently, all twenty-five constraints use the common feasibility test $v_j(\mathbf{x})\geq t_j(\mathbf{x})$. Equivalently, the residual
\begin{equation}
    d_j(\mathbf{x})=t_j(\mathbf{x})-v_j(\mathbf{x})
\end{equation}
is nonpositive for a satisfied constraint and positive for a violation. This common residual convention enables direct integration into optimization and machine-learning workflows.

Table~\ref{tab_2} reports the individual constraint-satisfaction rates in the random subset, which range from 8.6\% to 96.9\%, with a mean marginal satisfaction rate of 47.2\%. Although many random designs satisfy individual checks, each design must satisfy all twenty-five simultaneously. Across the 6,020 retained random designs, the mean number of violated constraints was 13.192, so the average design failed more than half of the checks. Moreover, the constraints are coupled because several depend on the same dimensions, thicknesses, and member spacings; their joint pass rate therefore cannot be estimated by multiplying the marginal pass rates. Empirically, no design in the retained random subset satisfied all twenty-five constraints. This cumulative requirement motivates the constraint-aware generation methods described next.

\section{Constraint-Aware Design Generation}
The absence of a fully constraint-feasible design among the randomly sampled structures motivates the development of methods that can guide candidate designs toward feasible regions of the design space. This section evaluates two such methods. The first is a surrogate-guided inverse-design method that uses neural-network predictions within a stochastic gradient Langevin dynamics (SGLD)-inspired sampling procedure. The second is an equation-informed parametric repair method developed with large language model (LLM) assistance. These proof-of-concept studies demonstrate uses of the MiDShip dataset and its geometry-generation and evaluation pipeline; neither is claimed as a novel optimization or repair algorithm.

\subsection{Surrogate-Guided Inverse Design of Ship Structures}

This proof-of-concept uses surrogate gradients to reduce predicted constraint penalties, while a lower-weighted secondary objective favors lower structural density and higher maximum bending moment. It demonstrates the workflow from surrogate fitting through CAD regeneration and constraint evaluation rather than a new prediction architecture or optimization algorithm. The experiment evaluates feasibility but does not separately test improvements in either structural geometric measure used in the secondary objective.

To enable this task, a neural network was trained to predict the twenty-five constraint residuals directly from the parametric design vectors. The generated SGLD candidates were evaluated using the implemented constraint calculations described in this paper and released with the dataset. Before training, each raw residual column was transformed according to
\begin{equation}
    \widetilde{d}_j(\mathbf{x})=
    \frac{2}{1+\exp\left[5d_j(\mathbf{x})/(d_{j,\min}+10^{-6})\right]}-1,
\end{equation}
where $d_{j,\min}$ is the most negative value of residual $j$ in the training set and $10^{-6}$ is a numerical stabilizer. This transformation maps the rule boundary to zero, violations to positive values, and increasingly satisfied designs to negative values. After the 30 CAD failures were excluded, the 6,020 retained random designs were divided into a ship-class-stratified 80\% training set and 20\% validation set using random seed 41. Input standardization and the residual transformations were fitted using only the 4,816 training designs. The constraint surrogate was a fully connected neural network with 120 input nodes, two hidden layers of 256 nodes each, and an output layer of 25 nodes, trained using mean squared error loss. Predictive performance was evaluated on all 1,204 validation designs using the coefficient of determination ($R^2$) and normalized root mean squared error ($NRMSE$), both calculated in the logistic-transformed residual space defined above. For each constraint output, $NRMSE$ was calculated by dividing the root mean squared error by the range (maximum minus minimum) of the corresponding logistic-transformed true residuals in the validation set. Table~\ref{tab_2} reports the validation results for each output. Across the twenty-five outputs, validation-set $R^2$ values ranged from 0.712 to 0.985 and $NRMSE$ values ranged from 3.8\% to 15.0\%.

The lowest validation $R^2$ values were for the transverse-bulkhead horizontal- and vertical-stiffener and deck transverse-stiffener section-modulus constraints. These transformed-space metrics do not measure errors in physical units or feasibility-boundary accuracy. No further investigation of these weaker fits was performed. Generated-design feasibility was determined using the implemented constraint calculations after geometry generation, rather than surrogate predictions alone.

A separate surrogate for the structural geometric measures used 120 input nodes, two hidden layers of 128 nodes each, and three outputs. It predicted the logarithms of structural density, cross-sectional area, and maximum bending moment. Before calculating the validation metrics, the predictions and validation targets were transformed back to their original physical units. The respective validation-set $R^2$ values were 0.984, 0.912, and 0.964, and the corresponding $NRMSE$ values were 1.5\%, 4.9\%, and 2.5\%. These results demonstrate that the dataset supports prediction of both constraint residuals and structural geometric measures from the parametric representation.

\begin{table*}[t]
\caption{Comparison of feasibility and constraint-violation statistics for the complete dataset subsets and the method-specific starting populations. The SGLD seed pool and repair source designs are drawn from the random subset and are not additional dataset instances. Feasibility is the fraction of designs satisfying all twenty-five implemented constraints, and the mean constraint-violation (CV) count is the total number of violations divided by the number of designs. Candidate parameter sets that failed CAD generation are excluded. MNN values are means $\pm 2s$ across equal-size 100-design sets, where $s$ is the sample standard deviation. The random and repaired results use thirty sets sampled without replacement within each set, with possible overlap across repeated draws; the SGLD result uses the five actual 100-candidate generation batches. MNN was not calculated for the method-specific starting populations. The MNN ranges describe between-set dispersion and are not confidence intervals. Full violation-count distributions are reported in Appendix Table~\ref{tab_6}. Arrows indicate the preferred direction for feasibility and CV count; MNN is reported descriptively, and boldface denotes the best feasibility and CV values.}
\label{tab_1}
\centering
\begin{tabular}{lcccc}
\toprule
Population & Designs & Feasibility rate ($\uparrow$) & Mean CV count ($\downarrow$) & MNN distance \\
\midrule
Random designs & 6,020 & 0.000 & 13.192 & $3.729 \pm 0.052$ \\
SGLD seed pool & 25 & 0.000 & 2.360 & -- \\
SGLD-generated designs & 496 & 0.649 & 0.409 & $1.144 \pm 0.104$ \\
Repair source designs & 3,127 & 0.000 & 10.288 & -- \\
Repaired designs & 6,237 & $\mathbf{0.794}$ & $\mathbf{0.296}$ & $3.495 \pm 0.133$ \\
\bottomrule
\end{tabular}
\end{table*}

\begin{table}[t]
\caption{Per-constraint satisfaction rates in the retained random subset and constraint-surrogate validation performance. Random satisfaction is the percentage of the 6,020 retained random designs that satisfy each constraint. The $R^2$ and normalized root mean squared error ($NRMSE$) values use all 1,204 designs in the class-stratified validation set and are calculated in logistic-transformed residual space. SM denotes section modulus.}
\label{tab_2}
\centering
\resizebox{\columnwidth}{!}{
\begin{tabular}{@{}lccc@{}}
\toprule
Constraint & \shortstack{Random satisfaction\\rate (\%)} & \shortstack{Validation\\$R^2$} & \shortstack{Validation\\$NRMSE$ (\%)} \\
\midrule
Double-bottom height & 83.3 & 0.931 & 6.3 \\
Bottom-floor thickness & 71.9 & 0.977 & 5.0 \\
Bottom-girder thickness & 13.6 & 0.967 & 5.1 \\
Bottom-floor spacing & 73.3 & 0.972 & 5.4 \\
Inner-bottom deck thickness & 8.9 & 0.891 & 8.3 \\
Hopper-plate thickness & 44.2 & 0.912 & 7.5 \\
Bottom long.-stiffener SM & 21.6 & 0.938 & 7.2 \\
Bottom trans.-stiffener SM & 8.6 & 0.868 & 8.0 \\
Hopper-stiffener SM & 44.8 & 0.911 & 8.4 \\
Side-frame SM & 10.4 & 0.879 & 8.2 \\
Web-frame SM & 30.1 & 0.973 & 5.3 \\
Web-frame depth & 40.1 & 0.981 & 4.8 \\
Web-frame thickness & 74.8 & 0.979 & 5.1 \\
Side-stringer SM & 25.9 & 0.958 & 6.2 \\
Side-stringer depth & 12.7 & 0.962 & 4.9 \\
Trans.-bulkhead thickness & 35.6 & 0.870 & 10.9 \\
Trans.-bulkhead horiz.-stiffener SM & 63.2 & 0.712 & 15.0 \\
Trans.-bulkhead vert.-stiffener SM & 75.3 & 0.777 & 14.1 \\
Deck trans.-stiffener SM & 80.0 & 0.719 & 13.9 \\
Deck-plate thickness & 16.6 & 0.934 & 7.1 \\
Deck-beam SM & 96.9 & 0.895 & 5.9 \\
Deck-beam depth & 96.8 & 0.901 & 7.2 \\
Deck-beam thickness & 78.3 & 0.985 & 3.8 \\
Side-shell thickness & 38.4 & 0.964 & 6.6 \\
Bottom-shell thickness & 35.3 & 0.957 & 6.6 \\
\bottomrule
\end{tabular}
}
\end{table}

The inverse-design stage leverages both the constraint-prediction network and the network predicting structural geometric measures, using gradient information to iteratively update design vectors. Because neural networks are continuous and differentiable functions, they support gradient-based sampling methods. In this work, new designs were generated using a modified, SGLD-inspired approach~\cite{welling2011bayesian,nguyen2020nanomaterial}, informed by the GenPFN sampling framework~\cite{bagazinski2026genpfn}. A rectified-linear-unit penalty is applied to each predicted transformed residual relative to a normalized satisfaction margin. The constraint penalty directs updates toward predictions that have not reached the specified margin. The secondary objective favors lower predicted structural density and higher predicted maximum bending moment. Stochastic noise promotes variation among candidate vectors. Because this procedure uses gradients of engineering surrogate models rather than the gradient of a Bayesian posterior log density, it is described as SGLD-inspired. No claim is made that the procedure samples from a Bayesian posterior or inherits the convergence guarantees of conventional SGLD.

The update rule for one inverse-design step is given in Equation~\ref{eq_8}.

\begin{equation}
\label{eq_8}
\begin{aligned}
\mathcal{L}_{\mathrm{cv}}(\mathbf{x})
&=\sum_{j=1}^{25}\operatorname{ReLU}\!\left(\theta_{\mathrm{cv},j}(\mathbf{x})-\tau\right),\\
\mathbf{x}_{t+1}
&=\mathbf{x}_{t}-\alpha\nabla_{\mathbf{x}}\mathcal{L}_{\mathrm{cv}}(\mathbf{x}_{t})\\
&\quad-\beta\nabla_{\mathbf{x}}\theta_{\mathrm{p}}(\mathbf{x}_{t})
+\sigma\boldsymbol{\epsilon}_{t},\\
&\quad \boldsymbol{\epsilon}_{t}\sim\mathcal{N}(\mathbf{0}_{120},\mathbf{I}_{120}).
\end{aligned}
\end{equation}

Here, $\mathbf{x}_t\in\mathbb{R}^{120}$ is a standardized parameter vector at iteration $t$; $\boldsymbol{\theta}_{\mathrm{cv}}:\mathbb{R}^{120}\rightarrow\mathbb{R}^{25}$ is the constraint surrogate; and $\operatorname{ReLU}(u)=\max(0,u)$ is applied componentwise before the twenty-five penalties are summed. The scalar $\theta_{\mathrm{p}}$ is the predicted standardized log structural density minus the predicted standardized log maximum bending moment. The vector $\boldsymbol{\epsilon}_t\in\mathbb{R}^{120}$ supplies independent standard-normal noise in the standardized coordinates, and $\mathbf{I}_{120}\in\mathbb{R}^{120\times120}$ is the corresponding identity matrix. The implementation uses $\alpha=0.1$, $\beta=0.01$, $\sigma=0.001$, and $\tau=-0.7$. The value $\tau=-0.7$ was selected heuristically as a normalized satisfaction margin: a constraint penalty remains active after its predicted transformed residual crosses zero and vanishes only when that prediction is at most $-0.7$. It is neither a raw engineering threshold nor a 70\% satisfaction probability. The relatively small value of $\beta$ makes this preference for predicted structural geometric measures secondary to the constraint penalty. The SGLD updates treat all parameters as continuous variables. Before CAD generation, every updated parameter was converted to its intended datatype: Boolean variables were rounded to the nearest binary value, integer-valued variables were rounded to the nearest integer, and each one-hot-encoded group was projected by assigning 1 to its largest component and 0 to the remaining components.

Sampling was initialized from the twenty-five random-subset design vectors that satisfied the largest numbers of constraints. Under the implemented evaluator used for the reported results, these seeds averaged 2.360 violations (standard deviation 0.700; median 2): three seeds had one violation, ten had two, and twelve had three. Each initial candidate vector was formed by interpolating randomly selected seed vectors before the SGLD updates were applied. Five batches of one hundred candidate designs were generated using one hundred gradient steps, for 500 attempted candidates. The candidate vectors were then converted into 3D geometries and evaluated using the implemented constraint calculations. Four candidates encountered CAD-generation failures, leaving 496 successfully generated and evaluated SGLD designs. Of these, 322 had no violations, 145 had one, and 29 had two, giving a feasibility rate of $322/496=64.9\%$ and a mean of 0.409 violations. Relative to the seed-pool mean, this is a reduction of 1.951 violations, or 82.7\%. Because the candidates were initialized from interpolated seed vectors, this is a population-level comparison rather than a paired seed-to-candidate result. Relative to the complete retained random-subset mean of 13.192, the SGLD mean is 96.9\% lower. Appendix Table~\ref{tab_6} reports the seed and generated-design distributions, while Table~\ref{tab_1} provides the broader population comparison and equal-size-set MNN results.

Starting from the selected near-feasible seeds, the complete SGLD-inspired procedure produced a retained candidate set with substantially fewer constraint violations than the full random subset. This result characterizes the complete procedure and does not isolate the contribution of the secondary objective term. Under the equal-size-set MNN protocol, mean nearest-neighbor spacing in the scaled parameter representation was lower for the SGLD batches (1.144) than for the random-design sets (3.729). Figure~\ref{fig_3}(b) provides a qualitative two-dimensional view consistent with this local-spacing comparison; it is not independent quantitative evidence of global coverage. The 64.9\% retained-subset feasibility rate contrasts with the absence of designs satisfying all twenty-five constraints in the random subset, although a considerable fraction of the SGLD-generated designs retained one or more violations. These results motivate the second method, which repairs selected random designs using explicit information from the constraint equations.

\subsection{LLM-Assisted, Constraint-Informed Parametric Repair}
The second demonstration uses equation-informed parametric repair developed through LLM-assisted code analysis. The resulting rules were executed and evaluated through the same geometry-generation and constraint-evaluation pipeline as the original dataset.

Following the parametric repair approaches of Jelovica, Cai, and Samanipour~\cite{jelovica2024improved,samanipour2020adaptive,cai2021adaptive,cai2023neural}, GPT-5.6 Sol assisted analysis of the constraint equations, evaluator, geometry generator, parameter ranges, and aligned data. This analysis mapped each constraint to the parameters influencing its value or threshold and supported implementation of the repair rules.

The rules were iteratively tested on Rhino-generated structures, with later iterations targeting violations that remained after geometry regeneration.

The resulting algorithm is rule-based rather than learned; the LLM was not used during design generation or evaluation. Only parameters associated with violated constraints were eligible for modification. Double-bottom height could change when required by a repair. All other principal hull dimensions, ship-type indicators, and hatch-opening parameters remained fixed. Depth and thickness were repaired from their calculated shortfalls, and section-modulus constraints used the nonlinear plate--stiffener equations. Stiffener spacing, height, and flange width were considered before increasing integer-valued thicknesses. Repairs could exceed the nominal random-sampling ranges to meet sampled target margins. Rhino then applied geometric coupling and alignment before the structures were evaluated using the implemented constraint calculations.

The following subsection describes how repair targets were sampled and how the repaired designs were evaluated.

\subsection{Sampling and Evaluation of Repaired Designs}

The goal of the repair study was to generate a constraint-feasible derivative subset from randomly generated source designs, not to estimate the out-of-sample generalization of a learned repair model. Because the procedure applies explicit constraint equations rather than fitting a predictive repair model, a conventional train--validation--test split was not used.

The repair experiment selected every valid random design with at most thirteen constraint violations in the records used for selection. This criterion yielded 3,127 source designs from the 6,020 retained random designs. The cutoff was chosen to target approximately the better-performing half of the random subset by constraint-violation count, providing alternative designs from sources that already satisfied many of the implemented constraints. In the selection records, these designs had one to thirteen violations, with a mean of 10.078. The reported before-and-after comparisons use evaluations of the same source designs under the implemented evaluator. In those evaluations, violation counts ranged from one to fourteen, with a mean of 10.288, standard deviation of 2.629, and median of eleven (Appendix Table~\ref{tab_6}). Two repaired variants were generated from each source design, resulting in 6,254 candidate designs. Each variant retained the hull form, ship type, hatch configuration, and parameters not associated with a violated constraint from its source design. The two variants differed in their independently sampled target margins for the violated constraints. Depth and thickness constraints were repaired to exceed their thresholds by 10\%-35\%, section-modulus constraints by 35\%-60\%, and spacing constraints by 20\%-30\%.

The random target margins were intended to broaden the sampled parameter and response values for downstream model training and improve the chance of satisfying constraints after geometry regeneration. Parameter changes can alter both constraint values and their thresholds. Larger margins were assigned to section-modulus constraints because their dependence on the repaired dimensions is nonlinear. These generation settings can increase material use; they were selected for dataset augmentation rather than weight minimization.

For constraint $j$, the repaired target was defined as
\begin{equation}
    t_j^{*} = t_j + \left|t_j\right|\epsilon_j,
\end{equation}
where $\epsilon_j$ was sampled independently according to the constraint family. The absolute value maintains the intended repair direction for the negatively signed maximum-spacing constraint. Candidate vectors were required to be unique; duplicate vectors were resampled using new target margins.

Of the 6,254 candidate designs, 6,237 were successfully generated and evaluated, while 17 encountered repeated CAD-generation errors. Across the 6,237 retained source--repair pairs, the source designs averaged 10.287 violations and the repaired designs averaged 0.296. The mean paired reduction was therefore 9.991 violations, or 97.1\%, and every successfully evaluated repaired design had fewer violations than its source. Relative to the complete retained random-subset mean of 13.192, the repaired-design mean is 97.8\% lower. A total of 4,952 designs were fully feasible, corresponding to an aggregate feasibility rate of 79.4\% among the successfully evaluated repaired designs. The remaining evaluated designs included 914 with one violation, 206 with two, 142 with three, and 23 with four. The paired comparison excludes the 17 candidates that failed CAD generation because no post-repair constraint evaluation was available for them.

Under the equal-size-set MNN protocol, mean MNN was 3.495 for the repaired-design sets, compared with 3.729 for the random-design sets and 1.144 for the SGLD batches. Thus, the repaired-design sets had greater mean nearest-neighbor spacing than the SGLD batches and slightly less than the random-design sets in this scaled representation. This metric does not by itself establish global design-space coverage or engineering realism. Figures~\ref{fig_3}(c) and~\ref{fig_3}(d) provide complementary qualitative views of the repair-candidate distribution at two plotting scales. Table~\ref{tab_1} compares full-subset feasibility and constraint-violation statistics and the equal-size-set MNN results.

The full MiDShip dataset contains 12,753 designs: 6,020 random designs, 496 SGLD-generated designs, and 6,237 repaired designs.

The repaired designs that retained constraint violations were concentrated in a small subset of constraints, as shown in Table~\ref{tab_3}. Hopper stiffener section modulus was the most common failure, occurring in 745 designs, followed by transverse-bulkhead thickness in 505 designs and horizontal transverse-bulkhead stiffener section modulus in 340 designs. Hopper-stiffener failures occurred in 560 containership and 185 bulk-carrier designs.

\begin{table*}[t]
    \centering
    \caption{Constraints remaining violated after repair and evaluation using the implemented constraint calculations. Percentages use the 6,237 successfully evaluated repaired designs as the denominator.}
    \label{tab_3}
    \resizebox{\textwidth}{!}{%
    \begin{tabular}{lrr}
        \toprule
        Constraint & \shortstack{Number of\\designs} & \shortstack{Frequency among\\evaluated repaired\\designs (\%)} \\
        \midrule
        Hopper stiffener section modulus
            & 745 & 11.94\% \\
        Transverse-bulkhead thickness
            & 505 & 8.10\% \\
        Horizontal transverse-bulkhead stiffener section modulus
            & 340 & 5.45\%\\
        Vertical transverse-bulkhead stiffener section modulus
            & 182 & 2.92\% \\
        Deck-plate thickness
            & 37 & 0.59\% \\
        Hopper-plate thickness
            & 25 & 0.40\% \\
        Deck transverse-stiffener section modulus
            & 7 & 0.11\% \\
        Inner-bottom deck thickness
            & 3 & 0.05\% \\
        \bottomrule
    
    \end{tabular}
    }
\end{table*}

The remaining violations identify dependencies that were not fully represented in the repair algorithm. For example, the transverse-bulkhead thickness requirement depends strongly on bulkhead-stiffener spacing, but the CAD-generation pipeline adjusts the stiffener parameters during post-processing to align them with other structural members. The parameter values used to propose a repair can therefore differ from the final geometric spacing used during evaluation. Similarly, hopper stiffeners are active for both containership and bulk-carrier designs but are subject to different ship-class-dependent requirements. These interactions explain why a parameter-level repair can remain infeasible after the geometry is regenerated and are discussed further in the Dataset Limitations subsection.

\section{Conclusion}
This work introduced MiDShip, a machine-learning-oriented benchmark of 12,753 cargo-hold structural designs: 6,020 random, 496 SGLD-generated, and 6,237 repaired designs. To the authors' knowledge, based on the literature reviewed through August 2026, it is the first publicly released, machine-learning-oriented benchmark to synchronize parametric designs, full CAD geometries and simplified geometries formatted for FEA meshing, engineering drawings and annotations, bills of materials, structural geometric measures, and classification-rule evaluations for cargo-hold structures. The dataset spans three ship classes and includes twenty-five ABS rule-based constraints. The dataset and accompanying software for geometry generation, CAD modification, and structural evaluation are released under the GNU General Public License version 3.0.

The SGLD-inspired study retained 496 of 500 candidates after four CAD failures. Of these, 322 (64.9\%) satisfied all twenty-five constraints. The full retained SGLD subset averaged 0.409 violations. This mean was 82.7\% below the 25-seed mean of 2.360 and 96.9\% below the random-subset mean of 13.192. Under the equal-size-set protocol, mean MNN was 1.144 for SGLD batches and 3.729 for random-design sets.

The equation-informed repair study retained 6,237 of 6,254 candidates after 17 CAD failures. Of these, 4,952 (79.4\%) satisfied all constraints. The full retained repaired subset averaged 0.296 violations. This mean was 97.1\% below the paired-source mean of 10.287 and 97.8\% below the random-subset mean of 13.192. Under the same protocol, repaired-design sets had mean MNN 3.495, compared with 1.144 for SGLD batches and 3.729 for random-design sets.

These proof-of-concept studies illustrate two uses of MiDShip for constraint-aware generation. The primary contribution is the synchronized dataset and reproducible geometry-generation and evaluation pipeline, not either generation method. MiDShip supports research linking parametric representations, CAD geometry, structural geometric measures, drawings, and rule-based evaluations, but satisfaction of its constraints does not establish complete ship feasibility or regulatory approval.

\subsection{Dataset Limitations}
\label{subsec:dataset-limitations}
MiDShip represents synthetic cargo-hold regions rather than complete or empirical ship structures and therefore omits global arrangements and vessel-level loading. Its prescribed ranges and simplified coupling rules should not be used to infer the prevalence of configurations in operational vessels. The designs use uniform member spacing, planar and predominantly orthogonal geometry, and stiffened flat rather than corrugated transverse bulkheads; they omit localized reinforcement and other complex forms. Real structures may also use nonuniform spacing to accommodate equipment, rounded transitions at connections, and openings for intersecting members or lightening. These details were simplified to limit the number of design parameters and the complexity of the representation. The code supports lightening holes, but they are not parametrically included in this release.

MiDShip supports variation within its prescribed parametric representation. Structural arrangements outside that representation require extensions to the generator. A generated design may differ from every design in the dataset without introducing a new structural concept; the present studies do not evaluate design novelty or creativity.

The generated designs have not been compared with real ship designs or independently optimized reference designs. Models trained on MiDShip may learn features of its prescribed representation and generation procedures, so their applicability to operational vessels requires separate evaluation. Real-design data would be valuable for tasks involving the omitted structural details and for assessing transfer to operational structures. MiDShip provides a public resource for initial model development within its represented structural family.

The twenty-five implemented constraints represent a limited set of rule-derived scantling evaluations and do not constitute a complete or ABS-vetted approval process. Important loading, global-strength, fatigue, buckling, and manufacturability evaluations remain outside the current pipeline. In addition, vessel length is fixed at 250~m for rule calculations that require full-vessel length. The stored $T$ field is unused in these evaluations; calculations requiring draft instead use the reference value $T_{\mathrm{rule}}=0.667D$. Satisfaction of all twenty-five constraints should therefore be interpreted only as feasibility under the implemented research constraints, not as evidence that a design is ready for construction or regulatory approval.

The two design-generation studies also have methodological limitations. The SGLD-inspired procedure was initialized using the twenty-five random-subset designs that satisfied the largest numbers of constraints (22--24 of 25), while the repair study used selected source designs containing between one and thirteen violations, inclusive. Consequently, the reported feasibility rates do not measure performance from arbitrary starting designs. The repair procedure primarily increases dimensions or changes spacing to satisfy selected constraints and does not yet optimize the resulting material weight or other structural performance objectives. Because the constraint equations informed both repair development and final constraint evaluation, geometry regeneration provides a check of the repaired parameters but not an independent validation of the underlying rule implementation. Matched optimizer and repair baselines, ablation studies, and method-level generalization experiments were outside the scope of these proof-of-concept demonstrations. Consequently, the results demonstrate the feasibility of the implemented workflows but do not establish superiority, sample efficiency, or state-of-the-art performance.

Finally, the LLM-assisted development process depends on the supplied code, model version, development context, and human-guided iteration. The resulting repair algorithm did not fully capture several CAD post-processing dependencies and ship-class-specific requirements. In particular, automatic stiffener-alignment operations and class-dependent hopper-stiffener constraints contributed to the remaining violations. The reported results therefore characterize the implemented repair algorithm rather than a general capability of LLMs to solve ship structural design problems.

\subsection{Future Work}
\label{subsec:future-work}
Future work will address automated meshing and finite-element analysis, additional loading conditions and classification requirements, joint optimization of feasibility and performance, less favorable starting designs, and matched optimization and repair baselines. Other extensions include richer structural features, additional ship classes, and multimodal tasks linking parameters, 3D geometry, and drawings.

MiDShip could also provide a common platform for studying the effects of new or revised structural requirements. Candidate rules could be implemented in the evaluation pipeline and applied across a shared set of designs to compare constraint satisfaction. Adapting designs to the revised requirements could then reveal changes in scantlings and steel mass. With suitable data from existing vessels, the same workflow could examine how their cargo-hold structures compare with proposed rules.

Evaluations spanning alternative rule formulations and parameter settings could also provide training data for models that help identify constants, multipliers, and other rule parameters. Such models could support calibration against specified safety objectives and independent structural evidence. These applications would extend the present study's use of selected existing ABS requirements. Together, these developments would move MiDShip toward system-level workflows that integrate geometry, physics, and regulatory constraints.

\section*{Datasheet for the MiDShip dataset}
The following subsections provide the datasheet for the MiDShip dataset~\cite{gebru2021datasheets}.

\subsection*{Motivation}
\textbf{Why was this dataset created?}
This dataset was created to provide a publicly released, machine-learning-oriented benchmark for cargo-hold structural design and evaluation. 

\textbf{Who created this dataset?}
The dataset was created by the MIT DeCoDE Lab.

\subsection*{Composition}
\textbf{What do the instances that comprise the dataset
represent?}

The dataset represents synthetically generated parametric geometries of cargo-hold regions in ships rather than surveyed or as-built vessels. This portion of a ship experiences the greatest structural loading.

\textbf{How many instances are there in total?}

There are 12,753 retained designs in total: 6,020 random designs, 496 SGLD-generated designs, and 6,237 repaired designs.

\textbf{What does each instance consist of?}

Each instance in the dataset comprises a parametric design vector for a ship structure, full 3D geometries and simplified geometries formatted for FEA meshing, a list of every structural element and its geometric properties, structural geometric measures and constraint evaluations, three 2D engineering drawings, and corresponding versions of those same three drawings with labeled bounding boxes and associated metadata. Thus, each retained design includes six drawing files in total.

\textbf{Is there any missing data?}

Every retained design includes all modalities described above. Candidate parameter sets that failed CAD generation were excluded rather than retained as instances with missing data.

\textbf{How is the data associated with each instance organized?}

Data are formatted as .csv files for tabular information, .igs and .3dm files for 3D geometry, and .png files for 2D drawings. Each instance is identified by an integer, and files are named for each unique structure.

\subsection*{Data Collection Process}
\textbf{How was the data collected?}

The original 6,020 retained designs were produced by uniformly sampling generator inputs over prescribed ranges and then applying the rounding, coupling, and ship-class-specific activation rules described above. Consequently, the final reported variables are not necessarily marginally uniform. These randomly generated vectors were used to procedurally generate the remaining information for each design. The SGLD-generated subset was generated from selected near-feasible random-design seeds, and the repaired subset contains repaired variants of selected random designs. Accordingly, the three subsets are related and should not be interpreted as statistically independent samples.

\textbf{Who was involved in the data collection process?}

The dataset was generated by researchers at MIT using automated pipelines for geometry generation and design evaluation.

\textbf{Over what time frame was the data collected?}

The parametric design code was written in 2025 and 2026. The final dataset was generated in 2026. 

\textbf{How was it decided which data to collect and which to exclude?}

Candidate parameter sets that failed CAD generation in \emph{Rhinoceros 3D} were excluded from the retained dataset and from subsequent geometry-based evaluations: 30 of 6,050 random candidates, 4 of 500 SGLD-inspired candidates, and 17 of 6,254 equation-informed repair candidates. The corresponding candidate-attempt and CAD-failure counts are reported separately in the findings.

\subsection*{Preprocessing, Cleaning, and Labeling}

\textbf{Was any preprocessing, cleaning, or labeling involved?}

During sampling, an automated data-cleaning process aligned selected structural elements within the 3D geometries generated from the parametric designs.

\textbf{How was this done and by whom?}
The preprocessing was performed automatically using a Python script.

\subsection*{Uses}
\textbf{For which engineering-design tasks is this dataset suitable?}

This dataset is suitable for designing and training evaluation models for ship structures.

\textbf{Has this dataset been used for any tasks already?}

Yes. A constraint surrogate was trained and evaluated, and models predicting constraints and structural geometric measures were used in the SGLD-inspired demonstration.

\textbf{What structural geometric measures are available for design studies using this dataset?}

Available measures include structural density, center of gravity, second moments of area, and maximum bending moment. 

\subsection*{Distribution}
\textbf{How is this dataset distributed?}

The MiDShip dataset is available through Hugging Face (\url{https://huggingface.co/datasets/DeCoDELab/MiDShip}), and the accompanying code is available through GitHub (\url{https://github.com/noahbagz/MiDShip_Dataset}). The software is released under the GNU General Public License version 3.0. The GitHub repository provides the release version, release date, software license, commit identifier, file manifest, maintenance information, and recommended citation.

\textbf{Are there any restrictions or licenses on its use?}

The MiDShip dataset and accompanying software are released under the GNU General Public License version 3.0, and are available exclusively for noncommercial research and educational purposes.

\textbf{How long will the dataset remain available?}

This dataset is publicly available and is intended to remain available indefinitely. 

\subsection*{Maintenance}

\textbf{Who is responsible for the maintenance of the dataset?}

The dataset is maintained by the authors through Hugging Face: \url{https://huggingface.co/datasets/DeCoDELab/MiDShip}.

\textbf{How can individuals submit corrections or updates?}

Individuals can submit corrections and updates via GitHub pull requests and issues.

\textbf{Is there a versioning system in place?}

Automatic versioning of the codebase is available on GitHub: \url{https://github.com/noahbagz/MiDShip_Dataset}.

\section*{Acknowledgements}
Disclaimer: This research was funded by the American Bureau of Shipping (ABS). The opinions, findings, conclusions, technical approach, analysis, calculations, and recommendations expressed herein, including any use, interpretation, or derivation of ABS Rule requirements or formulas, are solely those of the author(s) and have not been reviewed, validated, or endorsed by ABS. Nothing in this paper may be relied upon as a statement or interpretation of the ABS Rules or as a substitute for the ABS Rules as published by ABS, which govern in all cases. ABS makes no representation or warranty as to the accuracy or fitness for any purpose of the material herein and assumes no liability arising from its use.

The authors thank Myles Wortham for his assistance in developing the automated 2D drawing scripts. 

\clearpage

\onecolumn
\appendix
\numberwithin{equation}{section}

\section{Design Parameters}
\label{app:design-parameters}
Table~\ref{tab_4} lists the 120 parameter fields included in the MiDShip design vectors and their random-generation specifications.

Table~\ref{tab:parameter-glossary} defines the abbreviations, data-type labels, and symbols used in Table~\ref{tab_4}. The stored $T$ field is included in the released vectors but is unused by the geometry generator and rule evaluator; the evaluator's reference draft is defined separately as $T_{\mathrm{rule}}=0.667D$.

\begin{table}[!htbp]
\caption{Abbreviations, data-type labels, and symbols used in the design-parameter specifications.}
\label{tab:parameter-glossary}
\centering
\small
\begin{tabular}{@{}p{0.21\columnwidth}p{0.70\columnwidth}@{}}
\toprule
\textbf{Term} & \textbf{Meaning} \\
\midrule
\texttt{num} & Number \\
\texttt{inter.} & Intermediate \\
\texttt{trans.} & Transverse \\
\texttt{vert.} & Vertical \\
\texttt{long} & Longitudinal \\
\texttt{blkhd} & Bulkhead \\
\texttt{cont} & Containership \\
\texttt{cat} & Ship-category indicator \\
\texttt{h} & Member height \\
\texttt{t} & Thickness \\
\texttt{w} & Flange width \\
\texttt{torc} & Boolean bit selecting a T- or C-shaped stiffener cross-section \\
\texttt{bool} & Boolean bit \\
\texttt{int} & Integer \\
\texttt{one-hot} & One-hot categorical indicator \\
\texttt{frac.} & Fraction of the listed reference dimension \\
\midrule
$L_{3h}$ & Modeled three-hold length \\
$B$ & Beam \\
$T$ & Stored draft field \\
$D$ & Structural depth \\
$R_b$ & Bilge radius \\
$D_b$ & Double-bottom height \\
$L_1$ & Bracket-leg dimension \\
$L_2$ & Bracket-leg dimension \\
\bottomrule
\end{tabular}
\end{table}

The lower and upper limits in Table~\ref{tab_4} are nominal specifications for generating the random subset. Some parameters, including the principal dimensions, bilge radius, and bracket dimensions, are sampled through coupled ratios or ship-class-specific relationships. These specifications are not universal feasibility bounds: post-sampling adjustments can modify the stored random-design values, and the equation-informed repair procedure can produce parameter values outside the nominal random-sampling ranges.

\subsection*{Principal-Dimension and Ship-Class Coupling Rules}

The principal dimensions were generated sequentially rather than sampled independently from the marginal limits in Table~\ref{tab_4}. First, the modeled three-hold length was sampled as $L_{3h}\sim\mathcal{U}(40,80)$~m. A temporary full-ship length, used only during sampling and not included in the released parameter vector, was then defined as $L_{\mathrm{OA}}=s_L L_{3h}$, where $s_L\sim\mathcal{U}(3,4.5)$. The remaining dimensions were generated as
\begin{equation}
\begin{aligned}
B &= r_B L_{\mathrm{OA}},       & r_B   &\sim\mathcal{U}(0.10,0.20),\\
D &= r_D B,                     & r_D   &\sim\mathcal{U}(0.30,0.80),\\
T &= r_T D,                     & r_T   &\sim\mathcal{U}(0.25,0.667),\\
D_b &= 1000r_{D_b}B,            & r_{D_b}&\sim\mathcal{U}(0.05,0.0675),\\
R_b &= r_R D_b/1000,            & r_R   &\sim\mathcal{U}(0.75,1.00),
\end{aligned}
\end{equation}
where $B$, $D$, $T$, and $R_b$ are in meters and $D_b$ is in millimeters. The overhang was stored as a dimensionless fraction $l_{\mathrm{overhang}}\sim\mathcal{U}(0.03,0.05)$ of $L_{3h}$. The cleanup procedure also enforces $R_b\leq D_b/1000$. The stored $T$ field is retained as a sampled parameter but is unused by the geometry generator and rule evaluator; rule calculations requiring draft use $T_{\mathrm{rule}}=0.667D$.

Ship class was sampled uniformly from tanker, containership, and bulk carrier, then represented by the one-hot fields at indices 57--59. Class-dependent parameters were handled as follows:

\begin{itemize}
\item \textit{Tanker}: containership parameters 66--79 and bulk-carrier parameters 80--91 were set to zero. The floor- and deck-bracket leg dimensions at indices 60, 61, 63, and 64 were sampled uniformly from 500 to 1,000~mm.
\item \textit{Containership}: bulk-carrier parameters 80--91 and bracket-leg dimensions 60, 61, 63, and 64 were set to zero. Container-step parameters 66--79 remained active within the limits in Table~\ref{tab_4}.
\item \textit{Bulk carrier}: containership parameters 66--79 were set to zero. Hopper parameters 80--91 remained active, and bracket-leg dimensions 60, 61, 63, and 64 were sampled uniformly from 3,000 to 5,000~mm.
\end{itemize}

Bracket thicknesses at indices 62 and 65 were sampled as integers from 10 to 25~mm for all classes; for containerships, these stored values are inactive because the corresponding bracket-leg dimensions are zero. During geometry generation, active bracket dimensions were subsequently aligned with neighboring stiffeners or girders. Independently of ship class, the longitudinal-bulkhead indicator at index 105 was activated with probability 0.2; when inactive, fields 106--115 were set to zero. When the hatch-opening indicator at index 116 was inactive, fields 117--119 were set to zero. Inner-side-shell thicknesses sampled below 10~mm were converted to zero, indicating that no inner side shell was present; when an inner side shell was present, the web-frame flange-width field at index 50 was set to zero. Boolean, integer, and plate-thickness fields were finally rounded to their intended data types, and the ship-class fields were projected to a single active one-hot category.

\clearpage
\begingroup
\small
\begin{longtable}{lllll}
\caption{The 120 MiDShip parameter fields and their nominal random-sampling specifications, including units, lower limits (LL), and upper limits (UL).}\label{tab_4}\\
\toprule
\textbf{Index} & \textbf{Name} & \textbf{Units} & \textbf{LL} & \textbf{UL} \\
\midrule
\endfirsthead
\caption[]{The 120 MiDShip parameter fields and their nominal random-sampling specifications, including units, lower limits (LL), and upper limits (UL) (continued).}\\
\toprule
\textbf{Index} & \textbf{Name} & \textbf{Units} & \textbf{LL} & \textbf{UL} \\
\midrule
\endhead
\midrule
\multicolumn{5}{r}{Continued on next page.}\\
\endfoot
\bottomrule
\endlastfoot
0 & L\_3h & m & 40 & 80 \\
1 & B & m & 15 & 80 \\
2 & T & frac. D & 0.25 & 0.667 \\
3 & D & m & 10 & 40 \\
4 & R\_b & frac. Db & 0.75 & 1 \\
5 & l\_overhang & frac. L\_3h & 0.03 & 0.05 \\
6 & Db & mm & 800 & 3500 \\
7 & bottom shell thickness & mm & 10 & 35 \\
8 & side shell thickness & mm & 10 & 35 \\
9 & top deck thickness & mm & 15 & 35 \\
10 & inner bottom thickness & mm & 15 & 35 \\
11 & trans. bulkhead thickness & mm & 10 & 35 \\
12 & inner side shell thickness & mm & 6 & 20 \\
13 & shear strake thickness & mm & 15 & 35 \\
14 & shear strake height & mm & 1000 & 2500 \\
15 & web frames per hold & int & 3 & 8 \\
16 & num bottom girders & int & 3 & 11 \\
17 & num inter. bottom stiffeners & int & 3 & 6 \\
18 & num top girders & int & 3 & 11 \\
19 & num inter. deck stiffeners & int & 3 & 6 \\
20 & num side shell stiffeners & int & 15 & 30 \\
21 & num vert. bulkhead stiffeners & int & 12 & 24 \\
22 & num trans. bulkhead stiffeners & int & 12 & 24 \\
23 & bottom girder thickness & mm & 10 & 35 \\
24 & inter. bottom stiff h & mm & 120 & 400 \\
25 & inter. bottom stiff t & mm & 7 & 12 \\
26 & inter. bottom stiff w & mm & 14 & 200 \\
27 & inter. bottom stiff torc & bool & 0 & 1 \\
28 & trans. deck beam h & mm & 300 & 1000 \\
29 & trans. deck beam t & mm & 10 & 25 \\
30 & trans. deck beam w & mm & 150 & 400 \\
31 & trans. deck beam torc & bool & 0 & 1 \\
32 & deck girder h & mm & 500 & 1600 \\
33 & deck girder t & mm & 10 & 25 \\
34 & deck girder w & mm & 300 & 600 \\
35 & deck girder torc & bool & 0 & 1 \\
36 & deck stiff h & mm & 120 & 320 \\
37 & deck stiff t & mm & 7 & 12 \\
38 & deck stiff w & mm & 14 & 200 \\
39 & deck stiff torc & bool & 0 & 1 \\
40 & blkhd trans stiff h & mm & 120 & 320 \\
41 & blkhd trans stiff t & mm & 7 & 12 \\
42 & blkhd trans stiff w & mm & 14 & 200 \\
43 & blkhd trans stiff torc & bool & 0 & 1 \\
44 & blkhd vert stiff h & mm & 120 & 320 \\
45 & blkhd vert stiff t & mm & 7 & 12 \\
46 & blkhd vert stiff w & mm & 14 & 200 \\
47 & blkhd vert stiff torc & bool & 0 & 1 \\
48 & web frame h & mm & 1000 & 2500 \\
49 & web frame t & mm & 10 & 25 \\
50 & web frame w & mm & 300 & 600 \\
51 & web frame torc & bool & 0 & 1 \\
52 & floor t & mm & 10 & 30 \\
53 & side shell long stiff h & mm & 120 & 400 \\
54 & side shell long stiff t & mm & 7 & 12 \\
55 & side shell long stiff w & mm & 14 & 200 \\
56 & side shell long stiff torc & bool & 0 & 1 \\
57 & cat tanker & one-hot & 0 & 1 \\
58 & cat container & one-hot & 0 & 1 \\
59 & cat bulkcarrier & one-hot & 0 & 1 \\
60 & floor web bracket L1 & mm & 500 & 1000 \\
61 & floor web bracket L2 & mm & 500 & 1000 \\
62 & floor web bracket t & mm & 10 & 25 \\
63 & deck web bracket L1 & mm & 500 & 1000 \\
64 & deck web bracket L2 & mm & 500 & 1000 \\
65 & deck web bracket t & mm & 10 & 25 \\
66 & cont step height & mm & 2600 & 5500 \\
67 & cont step width & mm & 2500 & 5500 \\
68 & cont step deck plate thickness & mm & 10 & 35 \\
69 & cont step side plate thickness & mm & 10 & 35 \\
70 & num cont step deck stiff & int & 3 & 6 \\
71 & cont step deck stiff h & mm & 120 & 320 \\
72 & cont step deck stiff t & mm & 7 & 12 \\
73 & cont step deck stiff w & mm & 14 & 200 \\
74 & cont step deck stiff torc & bool & 0 & 1 \\
75 & num cont step side stiff & int & 5 & 10 \\
76 & cont step side stiff h & mm & 120 & 320 \\
77 & cont step side stiff t & mm & 7 & 12 \\
78 & cont step side stiff w & mm & 14 & 200 \\
79 & cont step side stiff torc & bool & 0 & 1 \\
80 & bulk car bottom hopper t & mm & 15 & 35 \\
81 & bulk car top hopper t & mm & 15 & 35 \\
82 & num bulk car bottom hopper stiff & int & 3 & 6 \\
83 & bulk car bottom hopper stiff h & mm & 120 & 320 \\
84 & bulk car bottom hopper stiff t & mm & 7 & 12 \\
85 & bulk car bottom hopper stiff w & mm & 14 & 200 \\
86 & bulk car bottom hopper stiff torc & bool & 0 & 1 \\
87 & num bulk car top hopper stiff & int & 3 & 6 \\
88 & bulk car top hopper stiff h & mm & 120 & 320 \\
89 & bulk car top hopper stiff t & mm & 7 & 12 \\
90 & bulk car top hopper stiff w & mm & 14 & 200 \\
91 & bulk car top hopper stiff torc & bool & 0 & 1 \\
92 & num inter. trans frames & int & 2 & 4 \\
93 & inter. bottom frame h & mm & 120 & 320 \\
94 & inter. bottom frame t & mm & 7 & 12 \\
95 & inter. bottom frame w & mm & 14 & 200 \\
96 & inter. bottom frame torc & bool & 0 & 1 \\
97 & inter. side frame h & mm & 120 & 320 \\
98 & inter. side frame t & mm & 7 & 12 \\
99 & inter. side frame w & mm & 14 & 200 \\
100 & inter. side frame torc & bool & 0 & 1 \\
101 & inter. deck frame h & mm & 120 & 320 \\
102 & inter. deck frame t & mm & 7 & 12 \\
103 & inter. deck frame w & mm & 14 & 200 \\
104 & inter. deck frame torc & bool & 0 & 1 \\
105 & longitudinal bulkhead bit & bool & 0 & 1 \\
106 & num longitudinal bulkheads & int & 1 & 3 \\
107 & long blkhd plate t & mm & 10 & 18 \\
108 & long blkhd long stiff h & mm & 120 & 320 \\
109 & long blkhd long stiff t & mm & 7 & 12 \\
110 & long blkhd long stiff w & mm & 100 & 200 \\
111 & long blkhd long stiff torc & bool & 0 & 1 \\
112 & long blkhd vert stiff h & mm & 120 & 320 \\
113 & long blkhd vert stiff t & mm & 7 & 12 \\
114 & long blkhd vert stiff w & mm & 100 & 200 \\
115 & long blkhd vert stiff torc & bool & 0 & 1 \\
116 & hatch openings bit & bool & 0 & 1 \\
117 & hatch opening L & frac. L\_3h & 0.1667 & 0.3 \\
118 & hatch opening W & frac. B & 0.25 & 0.8 \\
119 & hatch opening R & mm & 25 & 200 \\
\end{longtable}
\endgroup

\clearpage
\section{ABS Constraint Information}
\label{app:abs-constraint-information}
The constraints used to evaluate the cargo-hold geometries are listed in Table~\ref{tab_5}. The experimental constraints were formulated using the July 2024 ABS Marine Vessel Rules. The table lists 38 rule-to-output applications drawn from 22 cited clauses and consolidated into twenty-five evaluated constraint outputs. When multiple applicable minimum requirements contribute to the same output, the greatest required value is used as the governing threshold. Many additional ABS requirements exist but require a priori knowledge of structural loading from the system-level design of the ship~\cite{ABS_MVR}.
\begin{table}[H]
\caption{ABS rule-to-output applications used to formulate twenty-five constraint outputs for selected spacing, sizing, and stiffness requirements. For outputs governed by multiple applicable minimum requirements, the greatest required value is used.}
\label{tab_5}
\centering
\small
%\resizebox{\columnwidth}{!}{
\begin{tabular}{ll}
\hline
\textbf{Requirement Number} & \textbf{Structural Constraint} \\
\hline
ABS Part 3, Ch. 2, Sec. 4/3.1  & Minimum Double Bottom Height \\
ABS Part 3, Ch. 2, Sec. 4/1.2  &  Minimum Double Bottom Height \\
ABS Part 3, Ch. 2, Sec. 4/5.1  & Minimum Bottom Floor Thickness \\
ABS Part 3, Ch. 2, Sec. 4/5.1  & Minimum Bottom Girder Thickness \\
ABS Part 3, Ch. 2, Sec. 4/5.1  & Maximum Bottom Floor Spacing \\
ABS Part 3, Ch. 2, Sec. 4/9.2  & Minimum Inner Bottom Deck Thickness \\
ABS Part 3, Ch. 2, Sec. 4/9.5  & Minimum Inner Bottom Deck Thickness \\
ABS Part 3, Ch. 2, Sec. 10/3.1  & Minimum Inner Bottom Deck Thickness \\
ABS Part 3, Ch. 2, Sec. 3/5.17  & Minimum Inner Bottom Deck Thickness \\
ABS Part 3, Ch. 2, Sec. 4/9.2  & Minimum Hopper Plate Thickness \\
ABS Part 3, Ch. 2, Sec. 4/9.5  & Minimum Hopper Plate Thickness\\
ABS Part 3, Ch. 2, Sec. 10/3.1  & Minimum Hopper Plate Thickness \\
ABS Part 3, Ch. 2, Sec. 3/5.17  & Minimum Hopper Plate Thickness\\
ABS Part 3, Ch. 2, Sec. 10/3.3 & Minimum Long. Bottom Stiffener Section Modulus \\
ABS Part 3, Ch. 2, Sec. 10/3.3 & Minimum Trans. Bottom Stiffener Section Modulus \\
ABS Part 3, Ch. 2, Sec. 10/3.3 & Minimum Hopper Stiffener Section Modulus \\
ABS Part 3, Ch. 2, Sec. 5/3.1  & Minimum Frame Section Modulus \\
ABS Part 3, Ch. 2, Sec. 6/3.1  & Minimum Webframe Section Modulus \\
ABS Part 3, Ch. 2, Sec. 6/3.5  & Minimum Webframe Depth \\
ABS Part 3, Ch. 2, Sec. 6/3.5  & Minimum Webframe Thickness \\
ABS Part 3, Ch. 2, Sec. 6/5.1  & Minimum Side Stringer Section Modulus \\
ABS Part 3, Ch. 2, Sec. 6/5.3  & Minimum Side Stringer Depth \\
ABS Part 3, Ch. 2, Sec. 9/5.1  & Minimum Transverse Bulkhead Thickness \\
ABS Part 3, Ch. 2, Sec. 10/3.1 & Minimum Transverse Bulkhead Thickness \\
ABS Part 3, Ch. 2, Sec. 9/5.3  & Minimum Trans. Blkhd. Vert. Stiffener Section Modulus \\
ABS Part 3, Ch. 2, Sec. 10/3.3 & Minimum Trans. Blkhd Vert. Stiffener Section Modulus \\
ABS Part 3, Ch. 2, Sec. 9/5.3  & Minimum Trans. Blkhd. Horiz. Stiffener Section Modulus \\
ABS Part 3, Ch. 2, Sec. 10/3.3 & Minimum Trans. Blkhd. Horiz. Stiffener Section Modulus \\
ABS Part 3, Ch. 2, Sec. 7/3.1  & Minimum Deck Stiffener Section Modulus \\
ABS Part 3, Ch. 2, Sec. 3/5.1  & Minimum Deck Thickness \\
ABS Part 3, Ch. 2, Sec. 4/9.5  & Minimum Deck Thickness \\
ABS Part 3, Ch. 2, Sec. 3/5.17  & Minimum Deck Thickness \\
ABS Part 3, Ch. 2, Sec. 8/5.3  & Minimum Deck Beam Section Modulus \\
ABS Part 3, Ch. 2, Sec. 8/5.7  & Minimum Deck Beam Depth \\
ABS Part 3, Ch. 2, Sec. 8/5.7  & Minimum Deck Beam Thickness \\
ABS Part 3, Ch. 2, Sec. 2/3.9  & Minimum Side Shell Thickness \\
ABS Part 3, Ch. 2, Sec. 2/5.1  & Minimum Side Shell Thickness \\
ABS Part 3, Ch. 2, Sec. 2/3.15 & Minimum Bottom Shell Thickness \\
\hline
\end{tabular}
%}
\end{table}

\clearpage
\section{Source and Generated-Design Violation Distributions}
\label{app:violation-distributions}

\begin{table}[H]
\caption{Violation-count distributions for the method-specific source populations and retained generated designs. All counts use the implemented evaluator underlying the reported results. The SGLD comparison is population-level because candidates were initialized from interpolations of seed vectors. The repair-source column contains the 3,127 unique source designs; the paired source mean reported in the text is weighted over the 6,237 retained source--repair pairs and therefore differs slightly from the unique-source mean shown here.}
\label{tab_6}
\centering
\small
\begin{tabular}{ccccc}
\toprule
\shortstack{Violation\\count} & \shortstack{SGLD seeds\\($n=25$)} & \shortstack{Retained SGLD\\designs ($n=496$)} & \shortstack{Repair sources\\($n=3{,}127$)} & \shortstack{Retained repaired\\designs ($n=6{,}237$)} \\
\midrule
0  & 0  & 322 & 0   & 4,952 \\
1  & 3  & 145 & 3   & 914 \\
2  & 10 & 29  & 10  & 206 \\
3  & 12 & 0   & 19  & 142 \\
4  & 0  & 0   & 44  & 23 \\
5  & 0  & 0   & 103 & 0 \\
6  & 0  & 0   & 141 & 0 \\
7  & 0  & 0   & 204 & 0 \\
8  & 0  & 0   & 245 & 0 \\
9  & 0  & 0   & 308 & 0 \\
10 & 0  & 0   & 352 & 0 \\
11 & 0  & 0   & 437 & 0 \\
12 & 0  & 0   & 504 & 0 \\
13 & 0  & 0   & 578 & 0 \\
14 & 0  & 0   & 179 & 0 \\
\midrule
Mean & 2.360 & 0.409 & 10.288 & 0.296 \\
\bottomrule
\end{tabular}
\end{table}

\clearpage
\twocolumn
\bibliographystyle{unsrt}

\bibliography{paper}
\end{document}